\documentclass{article}

\usepackage{arxiv}
\usepackage{arxiv-more}

\usepackage[utf8]{inputenc} \usepackage[T1]{fontenc}    \usepackage{hyperref}       \usepackage{url}            \usepackage{booktabs}       \usepackage{amsfonts}       \usepackage{nicefrac}       \usepackage{microtype}      \usepackage{lipsum}
\usepackage{graphicx}
\graphicspath{ {./figures/} }
\usepackage{paracol}
\usepackage{graphicx}
\usepackage[utf8]{inputenc}
\usepackage{caption}
\usepackage{amssymb}
\usepackage{amsmath}
\usepackage{cite}
\usepackage{float}
\usepackage{subcaption}
\usepackage{lineno}

\usepackage{siunitx}
\usepackage{mathtools}

\usepackage{booktabs, multirow} \usepackage[table]{xcolor} 

\usepackage{pifont}

\def\citep{\cite}

\newcommand{\ttitle}{NLP-based classification of software tools for metagenomics sequencing data analysis into EDAM semantic annotation}

\newcommand{\tkeywords}{natural language processing, software tool classification, information retrieval, language models, metagenomics, EDAM ontology}

\def\tgithubURL{https://github.com/kaoutarDaoudHiri/NLP-based-classification-of-metagenomics-tools}

\def\tfigshareURL{https://doi.org/10.6084/m9.figshare.21350583}

\title{\ttitle}

\author{
Kaoutar Daoud Hiri \\
  Jo\v{z}ef Stefan International Postgraduate School\\
  Ljubljana, SI 1000, Slovenia\\
  BioSistemika\\
  Ljubljana, SI 1000, Slovenia\\
  \texttt{kdhiri@biosistemika.com} \\
\And
Matja\v{z} Hren \\
  BioSistemika\\
  Ljubljana, SI 1000, Slovenia\\
  \texttt{matjaz@scinote.net} \\
\And
Toma\v{z} Curk \\
  Faculty of Computer and Information Science\\
  University of Ljubljana\\
  Ve\v{c}na pot 113, 1000 Ljubljana, Slovenia\\
  \texttt{tomaz.curk@fri.uni-lj.si} \\
}

\date{}

\begin{document}
\maketitle

\begin{abstract}
\textbf{Motivation:} The rapid growth of metagenomics sequencing data makes metagenomics increasingly dependent on computational and statistical methods for fast and efficient analysis. Consequently, novel analysis tools for big-data metagenomics are constantly emerging. One of the biggest challenges for researchers occurs in the analysis planning stage: selecting the most suitable metagenomics software tool to gain valuable insights from sequencing data. The building process of data analysis pipelines is often laborious and time-consuming since it requires a deep and critical understanding of how to apply a particular tool to complete a specified metagenomics task. \\
\textbf{Results:} We have addressed this challenge by using machine learning methods to develop a classification system of metagenomics software tools into 13 classes (11 semantic annotations of EDAM and two virus-specific classes) based on the descriptions of the tools. We trained three classifiers (Naive Bayes, Logistic Regression, and Random Forest) using 15 text feature extraction techniques (TF-IDF, GloVe, BERT-based models, and others). The manually curated dataset includes 224 software tools and contains text from the abstract and the methods section of the tools' publications. The best classification performance, with an Area Under the Precision-Recall Curve score of 0.85, is achieved using Logistic regression, BioBERT for text embedding, and text from abstracts only. The proposed system provides accurate and unified identification of metagenomics data analysis tools and tasks, which is a crucial step in the construction of metagenomics data analysis pipelines.\\ \end{abstract}


\section{Introduction}

Metagenomics aims to provide insight into the genetic material present in various environmental samples. Viral metagenomics, for example, studies viral communities in water, soil, animals, and plants. The most common approach in metagenomics is to use high-throughput sequencing (HTS) of DNA or RNA, which generates millions of short-read nucleotide sequences. HTS data are used to detect and quantify genomes and transcriptomes in a biological sample.
The widespread adoption of HTS techniques in biological studies caused a rapid increase in the volume of metagenomics data that needs to be analyzed as efficiently and rapidly as possible. These metagenomics big data make the field increasingly dependent on computational and statistical methods that lead to discovering new knowledge from such data. Consequently, new analysis tools for big-data metagenomics are constantly emerging~\citep{Levin2018}, \emph{e.g.} 2500 new tools were produced in 2016.
HTS data analysis tools are computer programs that assist users with computational analyses of DNA and RNA sequences to understand their features and functionality using different analytical methods. Interest in such analysis may be motivated by different research questions, ranging from pathogen monitoring and identification to identifying all organisms in a sequenced biological sample. The standard approach to achieve this is to apply a combination of trimming, assembly, alignment and mapping, annotation, and other complex pipelines of software algorithms to HTS data.

HTS data analysis tools play an essential role in the pipeline construction process. Helping scientists select and use the appropriate tools facilitates the development of analysis-specific efficient pipelines and updating of existing ones. Individual institutions with various project constraints increasingly use metagenomics tools and gradually improve their knowledge and tool use. Under these circumstances, selecting the most suitable metagenomics software tool to gain valuable data insights can be complex and confusing for people involved in the pipeline-building process.

Before adding a tool to a pipeline, it is essential to know certain details about it. What are the required inputs? Which input and output file formats are supported? Most importantly, which data analysis task does the tool perform? ``Task'' refers to the function of the metagenomics tool or the analysis it performs. Having an overview of all the available tools for a given task is also crucial. The results provided by search engines are too unstructured to allow for a swift differentiation and comparison of similar tools. Furthermore, selecting a suitable tool for each data analysis step based on official publications and websites is not straightforward. Therefore, several benchmark studies tried to address ``the best tool for the task'' challenge, considering different perspectives, \emph{e.g.} plant-associated metagenome analysis tools \citep{Schilbert2020,Too2019,Behera2021}, machine learning-based approaches for metagenome analysis \citep{Jabeen2018,Too2019}, task-specific tools for mapping \citep{Schilbert2020,Hatem2013} and assembly \citep{Behera2021}, and complete pipelines for virus classification \citep{Nooij2018a,Jones2017a,Simmonds2018} and taxonomic classification \citep{Ye2019a,Larsen2014,Escobar-Zepeda2018}.

Other fields face a similar challenge with the abundance of software to classify. Machine learning approaches for software classification have been widely used in the cybersecurity domain~\citep{Agarkar2020, Halbouni2022}. Examples include data protection by developing misuse-based systems that detect malicious code and classify malware into different known families, \emph{e.g.} Worm, Trojan, Backdoor, Ransomware, and others. Another active area is anomaly-detection-based systems, which cluster binaries that behave similarly to identify new categories.

There is a plethora of metagenomics tool functions available. Understanding the functions of a given tool and comparing it with similar tools are complicated tasks. Different benchmark efforts for metagenomics tools are published regularly. Still, they are often incomplete, covering only a specific research question, including a limited set of tools, focusing extensively on technical metrics, or lacking transparency and continuity.

The \emph{Galaxy} platform~\citep{Afgan2018} provides a recommendation-based solution~\citep{Kumar2021} to help users create workflows. The recommendations are based on data from more than 18000 workflows and thousands of available tools for various scientific analyses. The deep learning-based recommendation system uses the tool sequences, the workflow quality, and the pattern analysis of tool usage to suggest highly relevant tools to the users for their specific data analysis. A set of tool sequences is extracted from each workflow created by the platform users. This approach is not fully personalized, as it only considers one metric, \emph{i.e.}, the similarity between tool sequences in workflows. The system will recommend the same next-step set of tools to all the users with the same built sequence. Furthermore, it limits the system to the workflow data available on the platform's internal database, where a certain type of analysis can predominate at a specific point in time. These constraints directly influence the quality of the recommendations, especially for minority user profiles, who will receive low-quality or unsuitable tool recommendations more frequently.

Machine learning-based classification systems of research papers were developed to help users find the appropriate paper. The search can be directed towards differentiating the topics~\citep{Kumar2016, Calvillo2016} or be focused on specific domains, \emph{e.g.} computer science~\citep{Mustafa2021, Kaur2015} or bioinformatics~\citep{Al-Mubaid2020}.

Classification systems use different algorithms and combinations of paper sections. In some works~\citep{Mustafa2021, Kyriakakis2019} they rely on established ontologies such as CSO - the computer science ontology~\citep{Salatino2018}, \emph{EDAM} - the ontology of bio-scientific data analysis and data management~\citep{Ison2013}, and SWO - the software ontology~\citep{Malone2014}.

We propose a machine learning-based system that uses curated and peer-reviewed abstract text descriptions to classify metagenomics tools into classes representing their main task. The classification system facilitates users to investigate tools quicker, decide where a tool fits in the metagenomics pipeline construction process, and quickly and efficiently select tools from 13 different classes.
\enlargethispage{12pt}
\begin{methods}
\section{Methods}
Our main goal was to be able to infer the main task of metagenomics tools from their description in natural text. We explored different combinations of the classification algorithm, its set of hyperparameters, the textual description, and the text embedding method to identify the best model for the task.

\subsection{Data sources}

The information contained in most scientific papers is typically divided into the title, abstract, introduction, methods, results, and discussion sections.
We manually gathered descriptions from the paper publications of 224 metagenomics tools. We collected the abstract sections in the ``abstracts only'' dataset and the methods section in the ``methods only'' dataset. We also prepared tool descriptions that include both the abstracts and methods sections in the ``abstracts+methods'' dataset (Supplementary Datasets~S1, S2, and S3 and see Supplementary Section~\ref{s:data}). All datasets include the title of the paper as the first sentence in the description of each tool. Each record in the collected datasets represents a single tool and contains the tool's name, description, and task (class) as represented in Table~\ref{tab:dataset_Excerpt}.

\begin{table}[!hb]
\processtable{Excerpt of raw ``abstracts only'' dataset for five tools belonging to different categories.\label{tab:dataset_Excerpt}}
{\begin{tabular}{@{}lllll@{}}\toprule 
Tool name & Tool description & Tool task (Class) \\\midrule
KrakenUniq & \vtop{\hbox{\strut KrakenUniq:confident and fast}\hbox{\strut metagenomics cl..}} & Classification\\
ViruDetect & \vtop{\hbox{\strut ViruDetect: An automated pipeline}\hbox{\strut or efficie..}} & Virus identification\\
ALLPATHS & \vtop{\hbox{\strut ALLPATHS: de novo assembly}\hbox{\strut of whole-genome sho..}} & Assembly\\
Bambino & \vtop{\hbox{\strut Bambino: a variant detector}\hbox{\strut and alignment view..}} & Visualisation\\
imGLAD & \vtop{\hbox{\strut imGLAD: accurate detection}\hbox{\strut and quantification}} & Abundance estimation\\\botrule
\end{tabular}}{}
\end{table}

\subsection{Task ontology}

The diverse and complex operations in bio-scientific data analysis lead us to rely on the well-established and comprehensive \emph{EDAM} ontology~\citep{Ison2013} to categorize the tools from a functional perspective. The 11 classes comprise bioinformatics operations and processes from the \emph{EDAM} ontology: ``(Sequence) alignment'', ``(Taxonomic) classification'', ``Mapping'', ``(Sequence) assembly'', ``(Sequence) trimming'', ``(Sequencing) quality control'', ``(Sequence) annotation'', ``(Sequence) assembly validation'', ``(RNA-seq quantification for) abundance estimation'', ``SNP-Discovery'', ``Visualization''. 

We defined two additional classes: ``Virus detection'' and ``Virus identification''. We assign to these two classes viral analysis tools classified as machine learning tools in \emph{EDAM} ontology, \emph{e.g.} DeepVirFinder~\citep{Ren2020} and VirNet~\citep{Abdelkareem2019}. We assign other viral analysis pipelines to the two classes even if the pipelines include several tools belonging to other \emph{EDAM} classes, such as K-mer counting, assembly, mapping, and others. Examples of such tools are Kodoja~\citep{Baizan-Edge2019}, VirFind~\citep{Ho2014} and VirusFinder~\citep{Wang2013}, which are all developed for virus detection and identification.

We assigned 224 tools into 13 tasks (classes). Some tools can be used for several tasks and thus belong to several classes. However, we only assigned them to one of the 13 classes, \emph{i.e.}, to the main task for which they were designed, see Supplementary Section~\ref{s:methods}. The obtained class distribution is shown in Figure~\ref{fig:class_distribution}.

\begin{figure}[htb]
\centerline{\includegraphics[width=7.5cm]{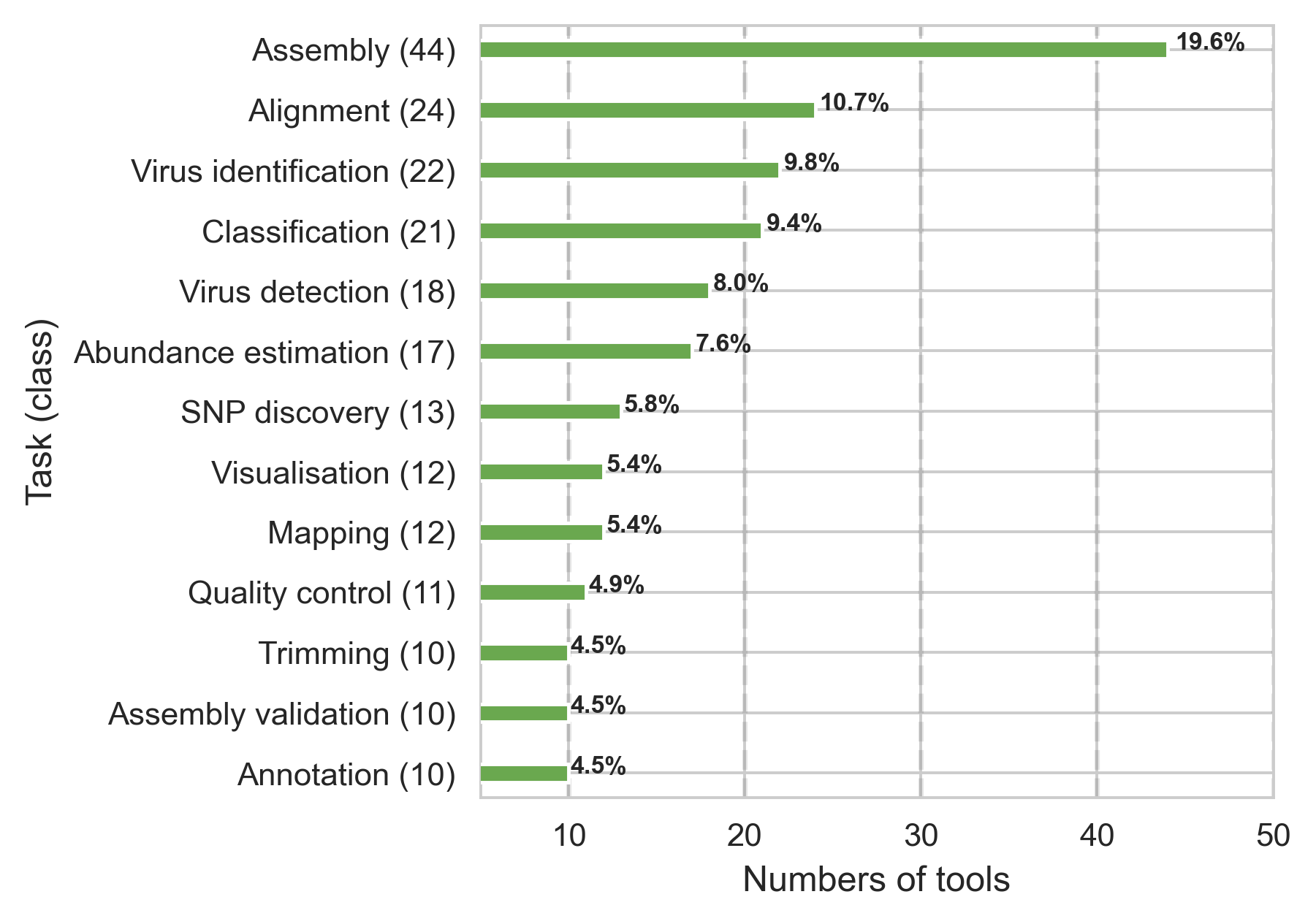}}
\caption{The class distribution shows the number of tools assigned to each task.}
\label{fig:class_distribution}
\end{figure}

\subsection{Data pre-processing}

\begin{figure}[t]
\centerline{\includegraphics[width=7.5cm]{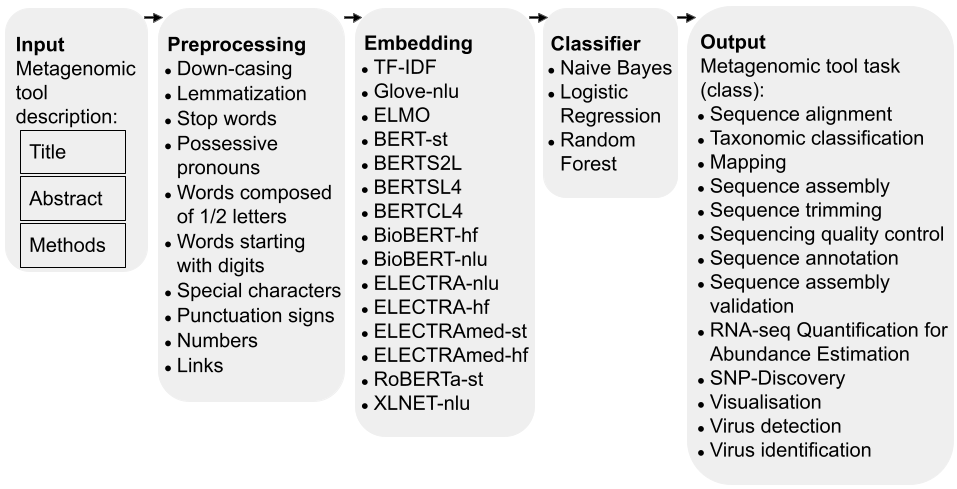}}
\caption{Process of extraction of information from text.}
\label{fig:information_Extraction}
\end{figure}

Before a classifier can use the available data, the appropriate pre-processing steps are required. The steps involved in extracting data from a tool description are summarized in Figure~\ref{fig:information_Extraction}. To create features from the raw text, train the classifiers and infer machine learning models, we performed the following steps: text cleaning and preparation, label coding, and vector representation of text (Supplementary Datasets~S4, S5, and S6). For text cleaning and preparation, we use downcasing, lemmatization, removal of stop words, possessive pronouns, words composed of one or two letters, words starting with digits, special characters, punctuation signs, numbers, and links. We represent the class variable as a nominal discrete variable with 13 different values. We then generated text vector representations, which are discussed in the following subsection.

\subsection{Vector representation of text}
To train the different classifiers, we represented the text description of the tools as a vector of numbers using language models, prediction-based and frequency-based techniques~(Supplementary Datasets~S7-S42).
\subsubsection{Word embedding methods}
We used and evaluated the 12 most commonly used approaches to extract features from the text. We describe them in the following paragraphs.

\textbf{TF-IDF} for a word in a document is calculated by multiplying the frequency of the term (term frequency)~\citep{Luhn1957} of a word in a document with the inverse document frequency of a word~\citep{Jones72astatistical} in a set of documents. If the word is very common and appears in many documents, this number will approach 0. Otherwise, the TF-IDF will approach 1.

\textbf{GloVe} Embeddings~\citep{pennington2014glove}, which stands for global vectors, capture the semantic context of words using both local statistics (local word context) and global statistics (word co-occurrences) to generate a word vector.
This regression neural network, trained on five combinations of general domain corpora (English Wikipedia and Gigaword), combines the advantages of global matrix factorization and local context window methods. It uses a gradient descent optimization algorithm and a decreasing weighting function where distant word pairs are expected to have less information about their relationship.

\textbf{ELMO}~\citep{peters2018}, deep contextualized word representation, represents each token based on the complete input sentence. The word representations combine the internal states of a pre-trained bidirectional language model (biLM) in a linear function learned by the end task model.

\textbf{BERT}~\citep{devlin2019bert}, which stands for Bidirectional Encoder Representations from Transformers, improves the fine-tuning-based strategies for applying pre-trained language representations to downstream tasks. It uses two unsupervised tasks during pre-training: binarized Next Sentence Prediction (NSP) and Masked Language Model (MLM).
Given a set of input tokens, the Masked Language Model randomly masks 15\% of the tokens. The goal is to predict the masked words based on their bidirectional context.
To understand the relationship between sentences, which is crucial for many downstream tasks, BERT pre-trains on NSP tasks which can be generated from the monolingual vocabulary. 
The final hidden state corresponding to the [CLS] token (the first token of every sequence) is used as the aggregate sequence representation for classification tasks.
In this work, we refer to L as the number of layers (transformer blocks), H as the number of hidden states, and A as the number of self-attention heads, and we report results on BERTBASE: L=12, H=768, A=12.
In addition to using the [CLS] token to represent a text sequence, we investigated three additional pooling strategies for BERTBASE, representing different choices of vectors from different layers:
\begin{itemize}
\item \textbf{BERTS2L}: Summing the vector embeddings generated from the Second to the Last Layer.
\item \textbf{BERTSL4}: Summing the vector embeddings generated from the Last Four Layers.
\item \textbf{BERTCL4}: Concatenation of the vector embeddings generated from the Last Four Layers.\vspace*{1pt}
\end{itemize}

\textbf{BioBERT}~\citep{devlin2019biobert} is a domain-specific language representation model based on the adaptation of BERT to the biomedical domain.
With the same architecture, weights, and Wordpiece vocabulary as BERT, BioBERT is pre-trained on corpora from the biomedical domain (PubMed abstracts and PMC full-text articles). BioBERT achieved a new state-of-the-art performance on three biomedical tasks: Biomedical named entity recognition (in terms of F1 score), biomedical relation extraction (in terms of F1 score), and biomedical question answering (in terms of mean reciprocal rank).

\textbf{XLNET}~\citep{yang2019xlnet} is a generalized AutoRegressive pre-training method that combines the best of AutoEncoding and Autoregressive language modeling while overcoming their limitations. XLNet is not based on a data corruption mechanism such as BERT. Consequently, special symbols used in pre-training are not missed in fine-tuning step. XLNET also improves the pre-training design architecture by (1) increasing the performance of long text-related tasks by including the segment recurrence mechanism and the relative encoding scheme of Transformer-XL in the training step, and (2) reparameterizing the Transformer-XL network to apply its architecture to permutation-based language modeling.

\textbf{RoBERTA}~\citep{liu2019roberta} is an optimized method of pre-training BERT-based models that demonstrate the benefits of bigger datasets, batches, and sequences to enhance model performance. The improved strategy also recommends training the models for a longer period, dynamically modifying the masking pattern used on the training data, and removing the next-sentence prediction objective. 

\textbf{ELECTRA}~\citep{clark2020electra} BERT is pre-trained using the masked language modeling approach to learn bidirectional word representations. ELECTRA (Efficiently Learning an Encoder that Classifies Token Replacements Accurately) proposes an alternative pre-training task (replaced token detection). The tokens are replaced with proposed alternatives produced by a generator network. Then the discriminator network predicts which token is original and which is a replacement.

\textbf{ELECTRAMed}~\citep{Miolo2021electramed}, based on ELECTRA, is a pre-trained domain-specific language model for the biomedical domain, inheriting the general-domain ELECTRA architecture learning framework and computational benefits.

\subsubsection{Short \emph{vs.} long text}
The complexity of the attention layer is quadratic to the length of the sequence~\citep{devlin2019bert}, therefore longer sequences are more expensive for BERT and BERT-based language models. The length of the text sequences cannot exceed 510 tokens, excluding special tokens ([CLS] and [SEP]). 
When analyzing the ``abstracts only'' dataset, we were not faced with this limitation. To extend the analysis to longer texts, we explored libraries \emph{NLU}~\citep{KOCAMAN2021100058}, \emph{sentence transformers} by UKP lab~\citep{reimers-2019-sentence-bert} and \emph{transformers} by Hugging Face~\citep{wolf2020huggingfaces}, depending on the availability of the models. We applied the long-text approach to all studied datasets, where we mapped input text into a fixed-length embedding based on the pre-trained model used. We also compared the performances of the direct, short-text and long-text approaches on the ``abstracts only'' dataset.

\begin{figure}[tb]
\centerline{\includegraphics[width=7.5cm]{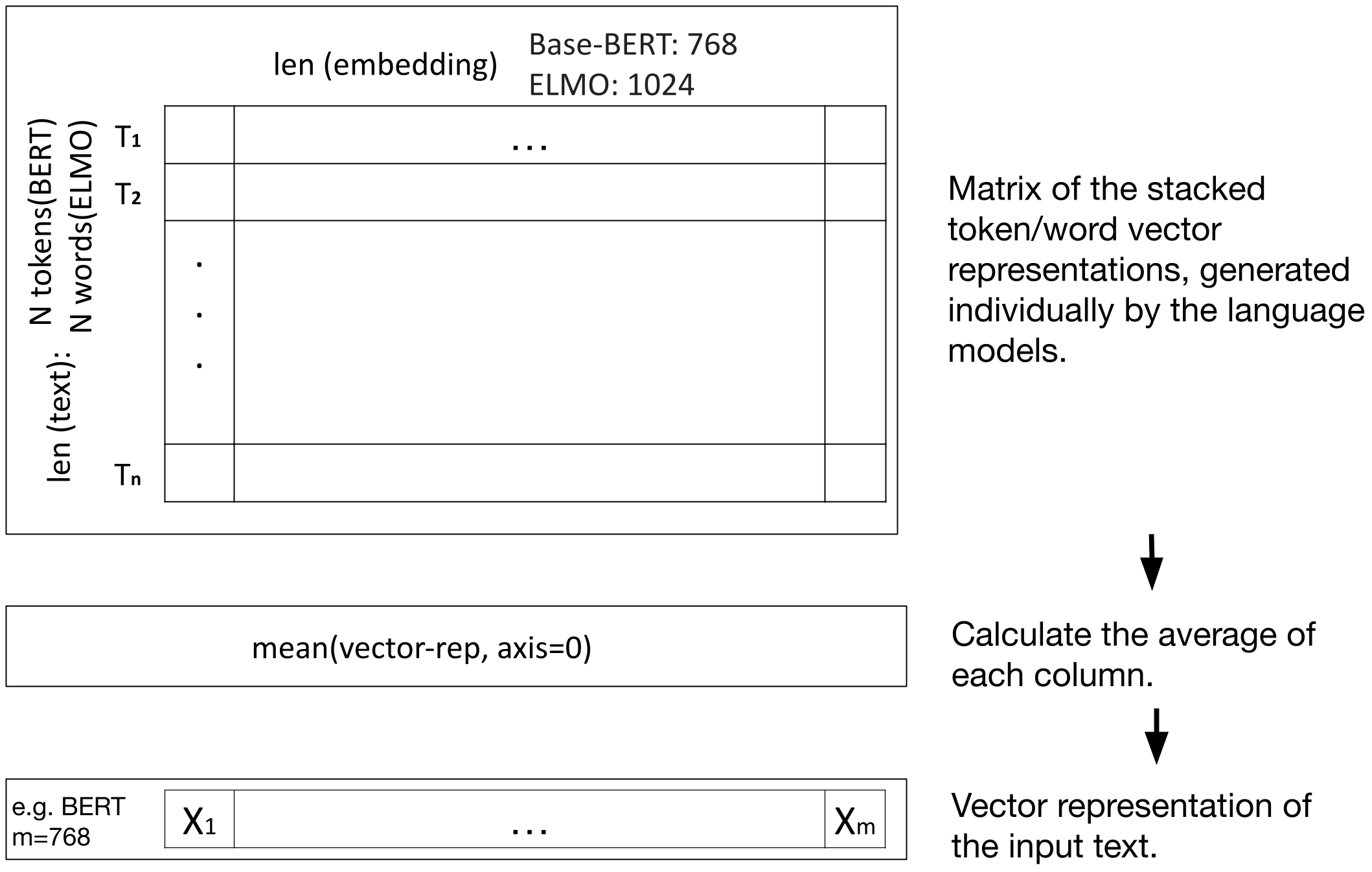}}
\caption{Method of generating the input text vector representation with the same length as the tokens/words vectors generated by a given embedding method.}\label{fig:Textembeddings_method}
\end{figure}

As shown in the Supplementary Table~\ref{stab:EmbeddingVectorSize}, the resulting word or token embeddings have different sizes, ranging from 100 to 3072 elements, depending on the algorithm we used to generate the vectors. Except for TF-IDF, all embedding methods were subjected to the following two steps to obtain the final sentence vector.
First, for each row in the dataset, we constructed an embedding matrix with $n$ rows and $m$ columns consisting of a list of words or tokens in the text and their corresponding numeric vector representations, as shown in Figure~\ref{fig:Textembeddings_method}, where $n$ is the number of words/tokens in the text description, and $m$ is the number of elements in the generated word embedding vectors.
Second, we calculated the average of the elements in each column of the resulting embedding matrix. Thus, we obtained sentence embeddings of the same size regardless of the length of the original text description.

\subsection{Learning algorithms}
To find which learning algorithm performed best on our data, we investigated three machine learning classification models with different parameter settings (Supplementary Data Table~S44): Logistic Regression (LR), Random Forest (RF), and Naive Bayes (NB).
We assembled a pipeline for the TF-IDF vectorizer and the classifiers so that they can be cross-validated together while setting different parameters. The feature extraction step was performed separately for text embedding models, and then we trained the classifiers on the resulting datasets.

\subsection{Model cross-validation on independent test set}
Firstly, we randomly split the dataset into a stratified train set and a test set, which we refer to as the independent test set. We then used repeated nested cross-validation on the training set with five outer folds and three inner folds. For hyperparameter tuning, we used scikit-learn~\citep{scikit-learn}'s \emph{GridSearchCV} function. The function was performed in the inner cross-validation on each inner training set and evaluated on the inner test set to select the best hyperparameter values. The model with the best hyperparameter setting was then trained on the outer train set and evaluated on the outer test set. For a list of hyperparameter values, see Supplementary Table~\ref{stab:CrossValidationHyperparameters}. Finally, the performance of the final model was estimated on the independent test set.
The methodology used to train each model was as follows:
\begin{itemize}
\item Assemble the pipeline.
\item Decide which hyperparameters we want to tune.
\item Build a grid containing the set of possible values of hyperparameters of the pipeline.
\item Define the metric to measure the performance of a model in a specific parameter setting. In our case, we use balanced accuracy.
\item Use a grid search cross-validation process to find the best combination of hyperparameters exhaustively, as described in the previous paragraph.
\item Obtain the model's performance with the best parameter settings using the Area Under the Receiver Operating Characteristic Curve.
\item Select the final model with the highest \emph{AUPRC} score in the five nested cross-validation rounds.
\item Obtain the final performance of the model on the independent test set using the different metrics described in the next section.\vspace*{1pt}
\end{itemize}
For TF-IDF, we fine-tuned its parameters in the inner loop~(Supplementary Data Table~S45).

Our datasets have unbalanced classes, making the classification task inherently more challenging, see Figure~\ref{fig:class_distribution}. As we wanted to detect the correct class for each tool, all target classes were equally important. The traditional classification algorithms we used in this study tend to favor majority over minority class elements due to their incorrect implicit assumption of an equal class representation during learning. To properly investigate the ability of our models to correctly detect each class, we used the F-score, the Area Under the Receiver Operating Characteristics curve (\emph{AUC-ROC}) score, and class-specific \emph{Precision}, \emph{Recall} and \emph{Accuracy}, as well as the Area Under the Precision-Recall Curve (\emph{AUC-PR}) score as a ranking metric, which is more suitable~\citep{Gaudreault2021} to assess the performance of the classifiers on unbalanced datasets.
To directly compare the models (classifier + embedding method) and select the best classification model while taking into account the class unbalance of the datasets, we measured \emph{Precision}, \emph{Recall}, and used the (\emph{AUC-PR}) score.
\end{methods}
\section{Results and Discussion}
We are reporting the results of the developed models. Each model consists of two steps: an embedding step and a classifier step, and these steps are performed using various methods and algorithms.
Firstly, we discuss the efficacy of various embedding and classification methods. Next, we identify the combination with the best predictive performance. Finally, we report and comment on the best model's typical misclassifications.

\subsection{BioBERT yields the best embedding representation}

When generated from the ``methods only'' dataset, TF-IDF embedding was by far the most informative embedding method for all three classifiers, with Random Forest classifier achieving the highest (\emph{AUC-PR}) score of 0.66, see Table~\ref{tab:AUPRC-scores}, Supplementary Figures~\ref{sfig:methodsPRCscores} and \ref{sfig:nemenyi-embeddings-3cx3d}. On the ``abstracts only'' and ``abstracts+methods'' datasets, TF-IDF was among the top five.

ELECTRA's implementation of the \emph{NLU} library was the least informative embedding in all datasets, see Supplementary Figure~\ref{sfig:LR+ELECTRA-NLU_methonly}. On one hand, the model is pre-trained on a general corpus, which produced the reverse effect of BioBERT embeddings on the classifiers' performance by reducing the informativeness of the model on a metagenomics-specific text (the tools' text descriptions). The slight increase in the classifiers' performance when trained on ELECTRA-hf embeddings instead of ELECTRAmed-hf also highlights the substantial effect of the pre-training corpus on the results. We deduce that a medical pre-training corpus does not improve the informativeness of the embeddings for metagenomics text, see the comparison of ELECTRA-hf and ELECTRAmed-hf in Supplementary Figure~\ref{sfig:abstractsPRCscores}. However, the most important factor influencing the low informativeness of ELECTRA-nlu embeddings is the pre-built pipelines used by the \emph{NLU} library. We can see a significant increase in the performance of the classifiers when the ELECTRA embeddings were generated without resorting to the \emph{NLU} library on the ``abstracts only'' dataset, see the comparison of ELECTRA-nlu and ELECTRA-hf in Supplementary Figure~\ref{sfig:abstractsPRCscores}.

On the ``abstracts only'' and ``abstracts+methods'' datasets, the set of BERT variations (BERTS2L, BERTSL4 and BERTCL4) which use different pooling strategies, in addition to the BERT-based methods such as BioBERT, built by pre-training BERT on the medical corpus, and the \emph{hugging face} implementation of ELECTRA which adapts the pre-training approach of BERT, consistently helped the classifiers yield better results.

\begin{table}[t]
\processtable{(\emph{AUC-PR}) scores of machine learning models by dataset. The embedding methods in bold were evaluated only on the ``abstracts only'' dataset. Rows are sorted in descending order of the best classifier (LR) performance on the ``abstracts only'' dataset. The highest scores in each column are in bold. Starred scores refer to the best-performing classifier for each dataset and embedding method pair.\label{tab:AUPRC-scores}}
{\begin{tabular}[t]{l@{\hspace{2.5\tabcolsep}}r@{\hspace{1\tabcolsep}}r@{\hspace{1\tabcolsep}}r@{\hspace{2.5\tabcolsep}}r@{\hspace{1\tabcolsep}}r@{\hspace{1\tabcolsep}}r@{\hspace{2.5\tabcolsep}}r
@{\hspace{1\tabcolsep}}r@{\hspace{1\tabcolsep}}r
}\toprule
&\multicolumn{3}{c}{abstracts only} &\multicolumn{3}{c}{methods only} &\multicolumn{3}{c}{abstracts+methods} \\\cmidrule{2-10}
&LR &RF &NB &LR &RF &NB &LR &RF &NB \\\midrule
\textbf{BioBert-hf} &\textbf{*0.85} &0.61 &0.56 &- &- &- &- &- &- \\
BERTS2L &*0.83 &0.47 &0.53 &*0.44 &0.31 &0.27 &\textbf{*0.84} &0.54 &0.53 \\
BERTSL4 &*0.81 &0.5 &0.47 &*0.4 &0.37 &0.27 &*0.83 &0.52 &0.49 \\
BERTCL4 &*0.79 &0.52 &0.49 &*0.41 &0.33 &0.24 &*0.82 &0.55 &0.54 \\
BERT-st &*0.75 &0.54 &0.53 &*0.48 &0.35 &0.3 &*0.73 &0.58 &0.54 \\
BioBERT-nlu &*0.73 &0.63 &0.49 &*0.46 &0.33 &0.28 &*0.77 &0.57 &0.48 \\
TFIDF &0.7 &\textbf{*0.8} &\textbf{0.71} &\textbf{0.62} &\textbf{*0.66} &\textbf{0.61} &0.73 &\textbf{*0.74} &\textbf{0.71} \\
\textbf{ELECTRA-hf} &*0.64 &0.5 &0.51 &- &- &- &- &- &- \\
\textbf{ELECTRAmed-hf} &*0.64 &0.49 &0.41 &- &- &- &- &- &- \\
RoBERTa-st &*0.61 &0.5 &0.43 &*0.38 &0.36 &0.31 &*0.59 &0.46 &0.4 \\
ELMO &*0.5 &0.46 &0.31 &*0.48 &0.38 &0.19 &*0.51 &0.48 &0.17 \\
XLNET-nu &*0.46 &0.33 &0.33 &*0.4 &0.32 &0.34 &0.43 &0.3 &*0.44 \\
GLOVE-nlu &0.33 &*0.44 &0.36 &0.2 &*0.36 &0.29 &0.22 &*0.35 &0.33 \\
ELECTRAmed-st &*0.24 &0.21 &0.15 &0.17 &*0.2 &0.13 &0.22 &*0.27 &0.16 \\
ELECTRA-nlu &*0.2 &0.17 &0.14 &0.12 &*0.18 &0.14 &0.2 &*0.22 &0.14 \\
\bottomrule
\end{tabular}}{}
\end{table}

\subsection{Logistic Regression outperforms Naive Bayes}

We evaluated the three classifiers (Logistic Regression, Random Forests, and Naive Bayes) with a statistical comparison using the Nemenyi test, as proposed by ~\citep{demsar2006statistical}, to reflect the overall performance of each classifier using their average ranks across the datasets. With $N$ classifiers, the method that performs the best has a rank of 1, while the method that performs the worst has a rank of $N$.

\begin{figure}[tpb]
\centerline{\includegraphics[width=5.8cm]{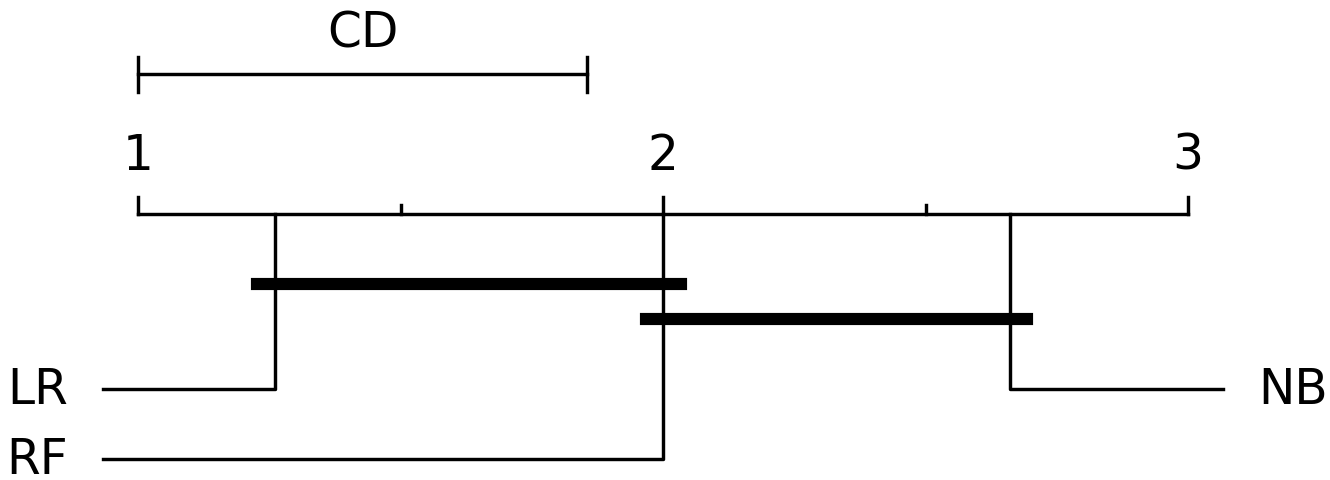}}
\caption{Comparison of classifiers with the Nemenyi test, tested on 15 embeddings on the ``abstracts only'' dataset. Groups of classifiers that are not significantly different are connected.}\label{fig:nemenyi-abst15}
\end{figure}

On the ``abstract only'' descriptions, the Logistic Regression classifier was significantly better than Random Forests and Naive Bayes. Logistic Regression outperformed the other two classifiers in every dataset, see Figure~\ref{fig:nemenyi-abst15} and Supplementary Figure~\ref{sfig:nemenyi-abst-methd-combined}. The mean (\emph{AUC-PR}) scores of Logistic Regression, Random Forest and Naive Bayes on ``abstracts only'' embeddings datasets are 0.60, 0.47 and 0.42 respectively; 0.37 , 0.34 and 0.28 on ``methods only'' embeddings datasets; and 0.57, 0.46 and 0.41 on ``abstracts+methods'' embeddings datasets. It is worth noticing that Logistic Regression consistently performed better than Random Forests and Naive Bayes on the ``abstracts only''-based embedding datasets, even if one among Naive Bayes or Random Forests was performing the worse.

The Logistic Regression significantly outperformed the other classifiers when trained on the language models-based embeddings of the three datasets. At the same time, the Naive Bayes always performed the worse through all the datasets and embedding strategies.

\subsection{Abstracts are crucial for good predictive performance}
All models showed lower performance when trained on the ``methods only'' dataset, compared to their performance on the ``abstracts only'' and the ``abstracts+methods'' datasets, see Supplementary Figure~\ref{sfig:methodsPRCscores} and Supplementary Figures~\ref{sfig:abstractsPRCscores} and~\ref{sfig:abstractsmethodsPRCscores}. The significant drop in (\emph{AUC-PR}) scores across all classifiers and embedding methods indicates that the text description of the tools covering only the methods section has poor informativeness. We note that the text informativeness of the methods sections in the scientific articles improved when we added the abstract section text, as shown in Figure~\ref{sfig:abstractsPRCscores}.

Our findings are consistent with previous studies where abstracts proved to be effective in capturing the context, \emph{e.g.} in bioinformatics research~\citep{Al-Mubaid2020} and computer science research papers~\citep{Kumar2016}.

\begin{figure}[t]
\centerline{\includegraphics[width=8.2cm]{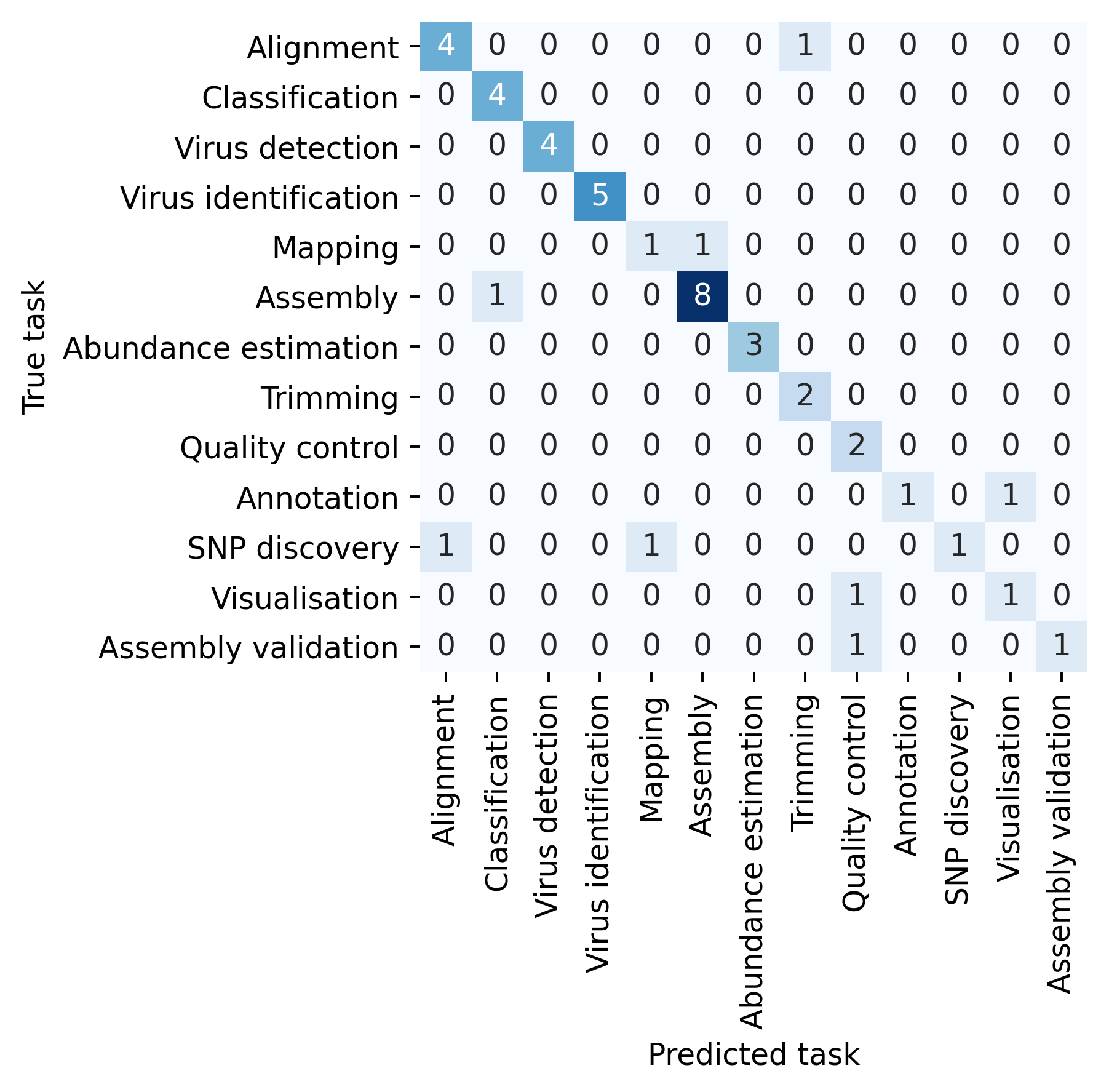}}
\caption{Logistic Regression on BioBert-hf - test set confusion matrix - ``abstracts only'' dataset.}\label{fig:LR+BioBert-hf_abstonly}
\end{figure}

\subsection{Evaluation of the best model}

The Logistic Regression classifier with \emph{BioBERT-hf} embedding strategy on the ``abstracts only'' dataset was the best model with a (\emph{AUC-PR}) score of 0.85, see Table~\ref{tab:AUPRC-scores}.
As shown in Figure~\ref{fig:LR+BioBert-hf_abstonly}, the model achieved perfect classification accuracy for the classes ``Classification'', ``Virus detection'', ``Virus Identification'', ``Abundance estimation'', ``Trimming'' and ``Quality control'', without completely missing any other class. Most of the other models had difficulty distinguishing ``Virus detection'' from ``Virus identification'' instances and ``classification'' from ``Abundance estimation'' or ``Assembly'', as shown in Supplementary Figures~\ref{sfig:LR+BERTS2L_abstmethd}, ~\ref{sfig:LR+BERTS2L_abstonly} and ~\ref{sfig:LR+BERTSL4_abstmethd}, which can be explained with the significant similarity between the three classes.
Supplementary Figure~\ref{sfig:PRCCurves} also highlights the improvement in the performance of the top models compared to the baseline (Supplementary Section~\ref{s:baseline}) and to the worst trained model (Logistic Regression classifier with ELECTRA-nlu embedding strategy on the ``methods only'' dataset).

The correct class of misclassified instances (Supplementary Table~\ref{stab:IndependentTestSetClassifications}) was usually in the top three classification probabilities predicted by the model (Supplementary Data Table~S46) for predicted probabilities. This pattern excluded two instances, ``Tagdb'' and ``DeepVariant'', where the correct class probabilities were ranked 7 and 8 out of 13 consecutively.
In addition, our model produced few seeming misclassifications when the tools in question performed more than one task:
\begin{itemize}
\item ``Savant''~\citep{Fiume2010} was misclassified as an ``Annotation'' tool instead of a ``Visualisation'' tool since it is a sequence annotation, visualization, and analysis framework, see Supplementary Figure~\ref{sfig:Savant}.
\item ``Kart''~\citep{Lin2017} was misclassified as a ``Trimming'' tool instead of an ``Alignment'' tool, as it performs a trimming step first by dividing long reads into shorter bits to align them independently, see Supplementary Figure~\ref{sfig:Kart}.
\item ``dnAQET''~\citep{Yavas2019} was misclassified as a ``Quality Control'' tool instead of an ``Assembly validation'' tool, as it requires calculating the contigs individual quality scores, see Supplementary Figure~\ref{sfig:dnAQET}.\vspace*{1pt}
\end{itemize}

This behavior highlights that the classifiers will always struggle to detect the correct class of a given metagenomics tool in this particular evaluation scheme where we assume a multi-class classification problem.

\section{Conclusion}
According to our results, the model composed of the Logistic Regression classifier trained on text representation of the ``abstracts only'' dataset generated using the Hugging Face implementation of \emph{BioBERT} has the best ability to distinguish between the different classes (Supplementary Data Table~S46). The considerable number of tools belonging to more than one class explains a subset of the misclassifications of the models as we picked only one class for each tool during manual data curation. A multi-label classification problem setting would address this issue.

Another reason for the misclassifications of the models is the absence of a controlled vocabulary or unified terminology in the metagenomics domain. We can see the interchangeable use of terms that refer to different processes for different people, \emph{e.g.} terms ``mapping'' and ``alignment'' have been generating a lot of confusion recently. In reality, both terms pertain to aligning sequences; however, whereas ``mapping'' aligns short sequences against a reference genome, ``alignment'' aligns short sequences against one another. The readers become confused when such terms are used as synonyms. Wider adoption of a unified vocabulary \emph{e.g.} \emph{EDAM}, would reduce inconsistencies in word usage and improve machine and human comprehension.

In our study, the abstract section was the most informative for classifying tools, as it summarizes the research paper well. In the case of the ``methods only'' dataset, the performance drop and inexplicable misclassifications on our independent test set may be caused by the noise introduced in the text descriptions and our manual text extraction process. 

The low informativeness of the methods section was previously reported for computer science research~\citep{Kumar2016} and some other research fields, \emph{e.g.} astrophysics research, where the Discussion section captures the most context. Identifying the most informative part of the paper, which describes a tool in detail, can be challenging. The description depends on the author's writing style and journal structuring methods, even if one looks at papers in the same research field.
Developing a novel text segmentation approach that can automatically detect and extract specific topics and sentences, depending on the query, might help improve information retrieval from metagenomics-specific text content.

We also note the significant influence of the text corpus used during the training phase of the embedding models, mainly when applied to domain-specific text data. The metagenomics field can primarily benefit from a language model, such as BERT, pre-trained on the metagenomics corpus.

Different embedding-pooling strategies also influence the performance of classifiers. Embedding models are pre-trained on a few specific Natural Language Processing (\emph{NLP}) tasks; with each transitioning step from one layer to another, the encoded information about a given word and its context becomes more relevant to these specific tasks. Therefore, the efficiency of such contextualized vector representations is application-specific. It also depends on the downstream task~\citep{Wang2019}, the linguistic properties of the datasets, and inference tools (Bi-LSTM, CNN or other).

We developed and tested several (\emph{NLP})-based classification models on a dataset of 224 manually collected tool papers, categorized into 13 different \emph{EDAM} classes (tasks). The \emph{Galaxy} platform database, for instance, contains more than 7800 tools corresponding to more than 40 classes (\emph{EDAM} operations). This fact highlights the need for a more generalized and human-independent system that can automatically look for newly published tools, scan their papers, cluster them to identify new categories and label them according to their classes.
We propose a classifier system that uses Logistic Regression trained on the \emph{BioBERT} embeddings of the tools' publications abstracts to map the tools' descriptions to a space of \emph{EDAM} classes(tasks). Oriented toward learning the ``essentials'' (the correct task a given tool can perform), it avoids its users' misidentification of tools characteristics and function. This system can be easily scaled by training on larger datasets with more classes simply by collecting scientific articles from online sources. It can also be integrated into platforms such \emph{Galaxy} to accelerate the integration of new tools to its internal database.

\section*{Funding}

This work is supported by the European Union's Horizon2020 research and innovation program under the Marie Sklodowska-Curie [grant agreement number 813542], project INEXTVIR, a Marie Sklodowska-Curie Innovative Training Network.

\section*{Availability}

Supporting data and code are available on Github \url{\tgithubURL}~and on Figshare \url{\tfigshareURL}.

\bibliographystyle{unsrt}

\bibliography{lit.bib}

\clearpage
\section*{Supplementary material}
\appendix
\counterwithin{figure}{section}
\counterwithin{table}{section}

\renewcommand{\thefigure}{S\arabic{figure}}
\renewcommand{\thetable}{S\arabic{table}}
\renewcommand{\thesection}{S\arabic{section}}
\renewcommand{\figurename}{Supplementary Figure}
\renewcommand{\tablename}{Supplementary Table}

\section{Preparation of datasets} \label{s:data}
In this section, we list the datasets used in the study. We prepared the following datasets by manually extracting sections from the tools' published articles:
\begin{itemize}
    \item abstracts stored as ``abstracts only'' dataset, 
    \item methods stored as ``methods only'' dataset, and 
    \item both sections stored as ``abstracts+methods'' dataset. 
\end{itemize}
We describe a collection of six datasets: 
\begin{itemize}
    \item three datasets of the tools' descriptions in their unprocessed raw form and
    \item three pre-processed tool descriptions datasets. 
\end{itemize}

\begin{figure}[H]
\subfloat[]{\includegraphics[width=0.5\linewidth]{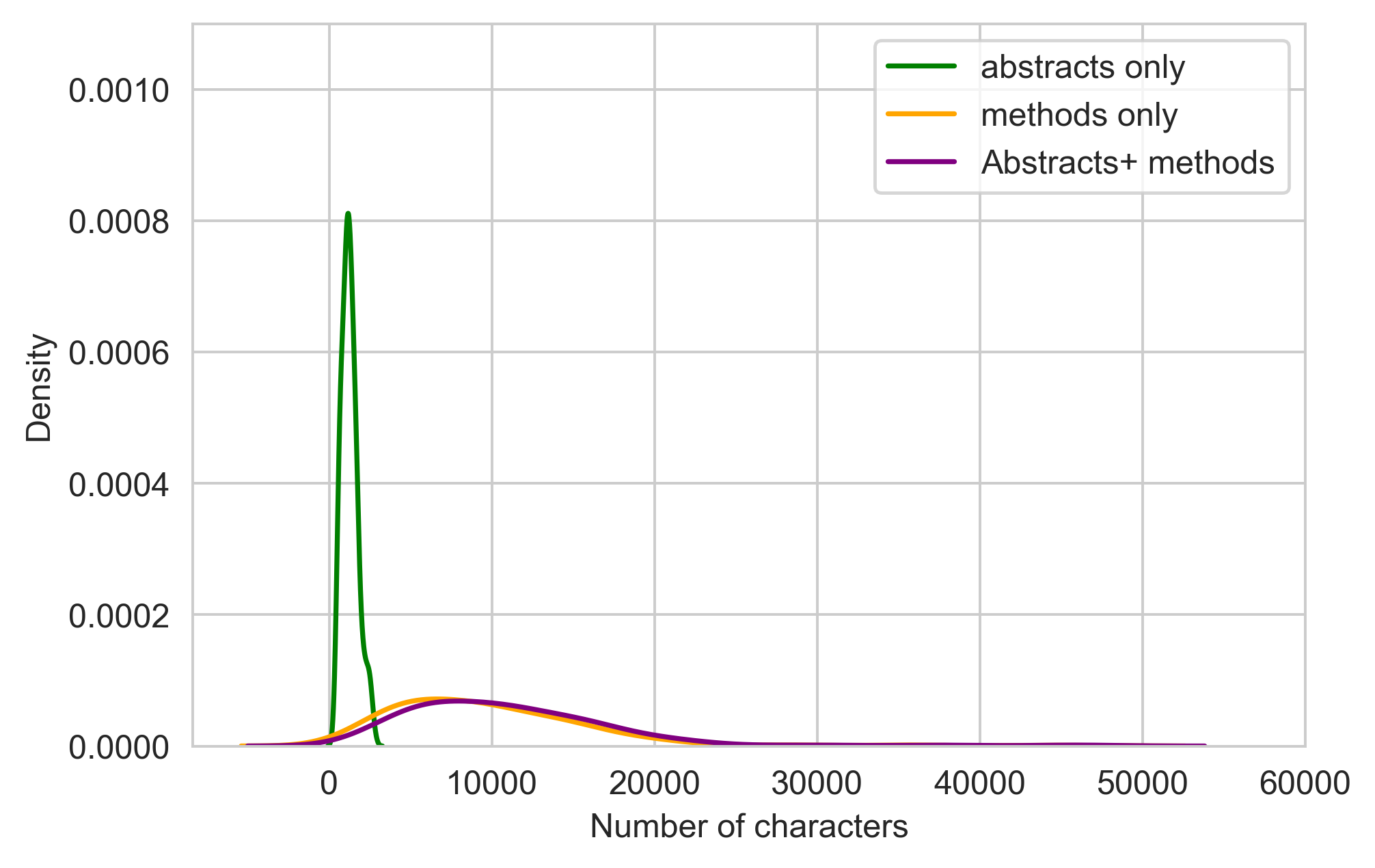}}
\subfloat[]{\includegraphics[width=0.5\linewidth]{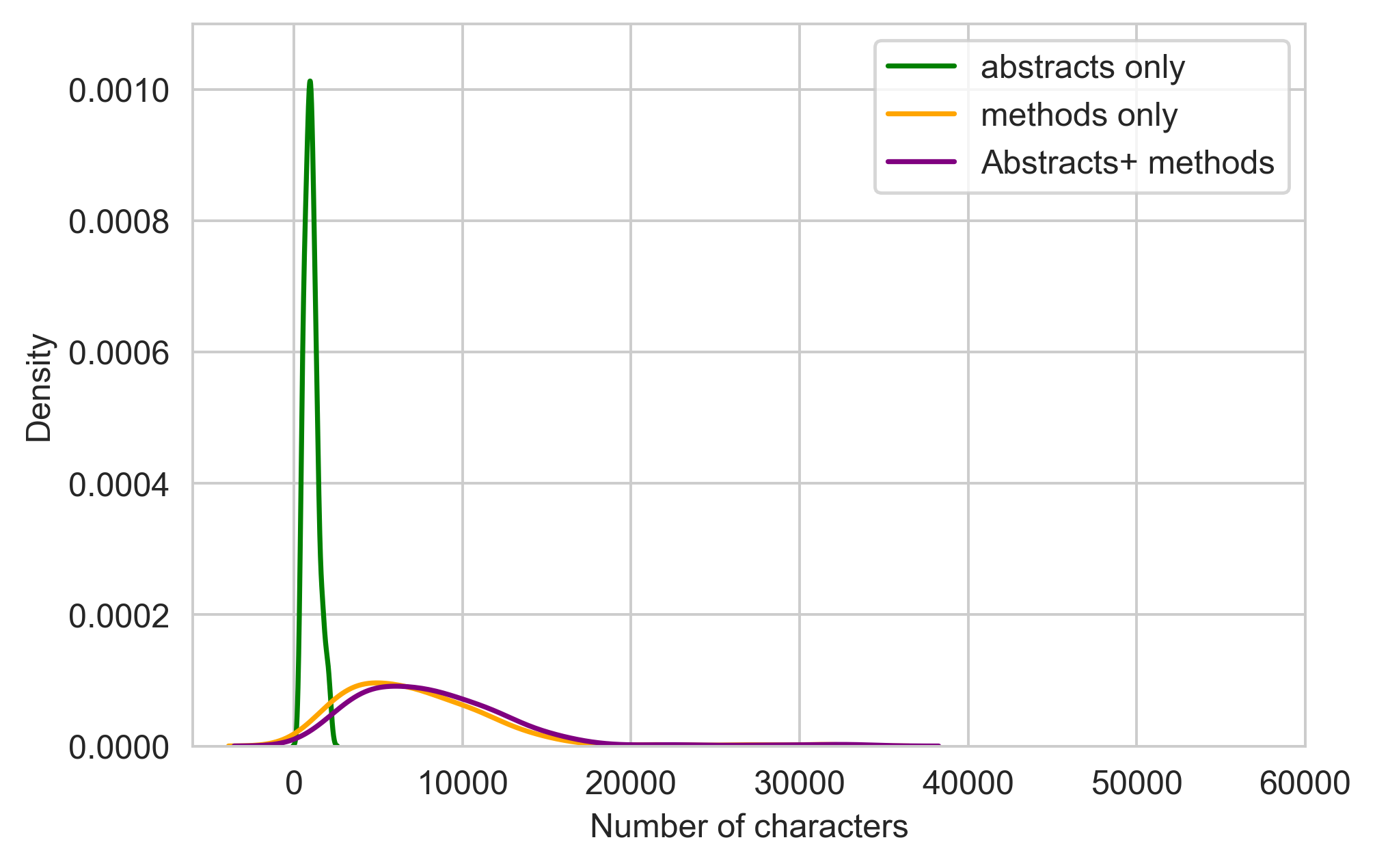}}\\
\caption{Distribution of the tool descriptions length in terms of the number of characters by dataset, on the (a) unprocessed text and (b) pre-processed text.}
\label{sfig:processedVSunprocessed_textsize_distribution_characters}
\end{figure}

\begin{figure}[!htb]
\centering
\subfloat[]{\includegraphics[width=0.4\linewidth]{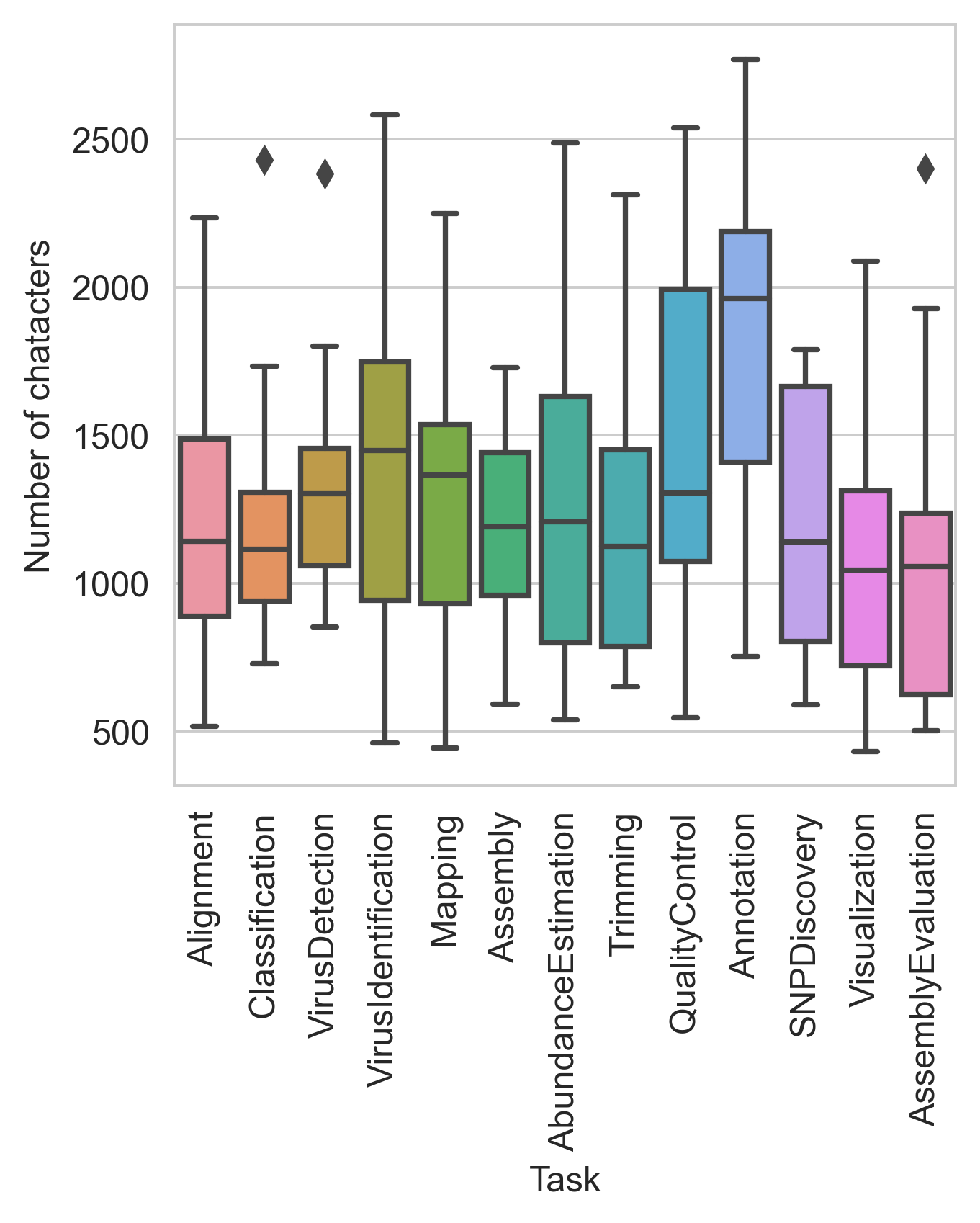}}
\subfloat[]{\includegraphics[width=0.4\linewidth]{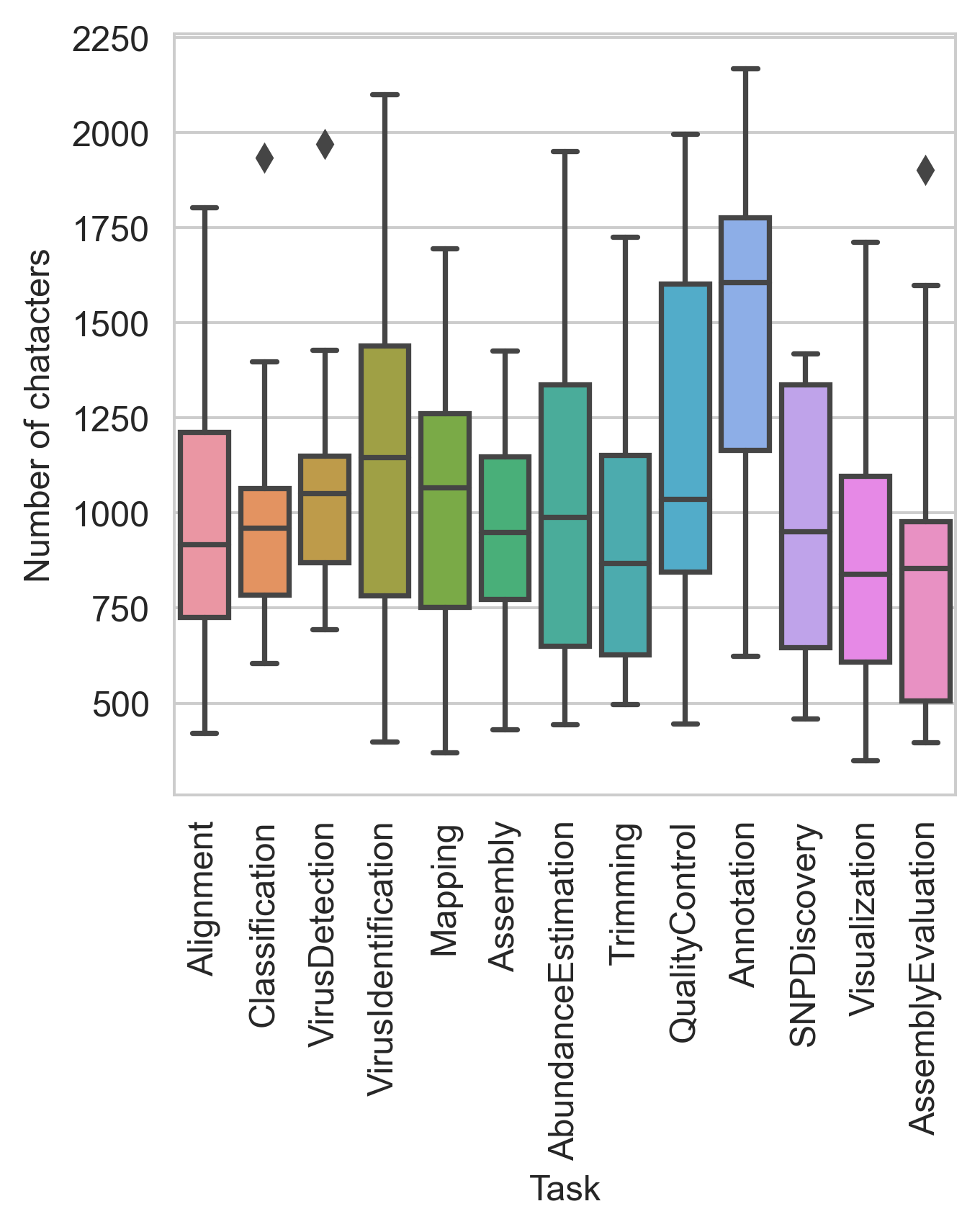}}\\
\caption{Distribution of the tool descriptions length in terms of the number of characters by task/category, (a) unprocessed text, (b) pre-processed text - ``abstracts only'' dataset.}\label{sfig:processedVSunprocessed_A_textsizeC_distribution_byclass}
\end{figure}

\begin{figure}[!htb]
\centering
\subfloat[]{\includegraphics[width=0.4\linewidth]{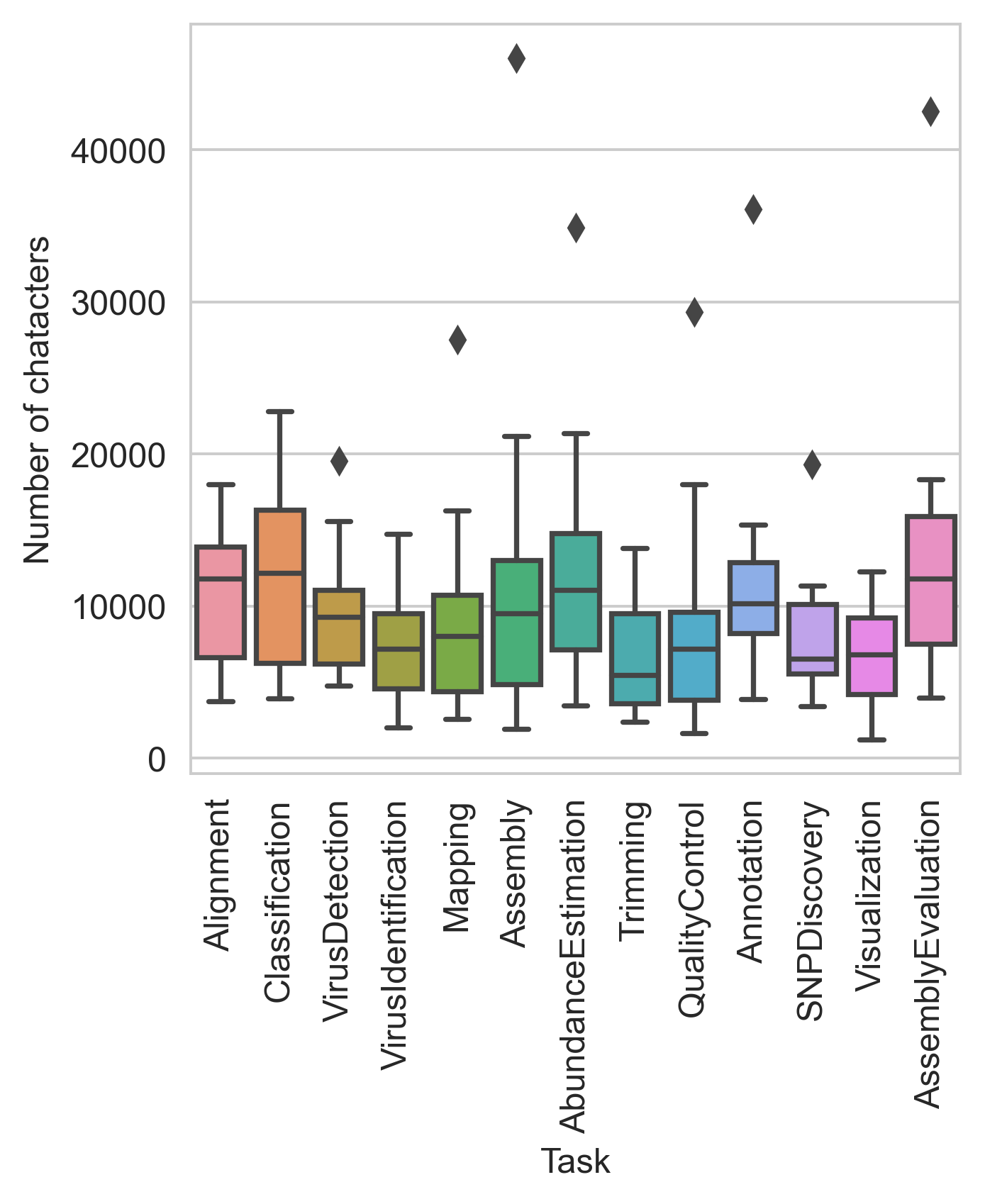}}
\subfloat[]{\includegraphics[width=0.4\linewidth]{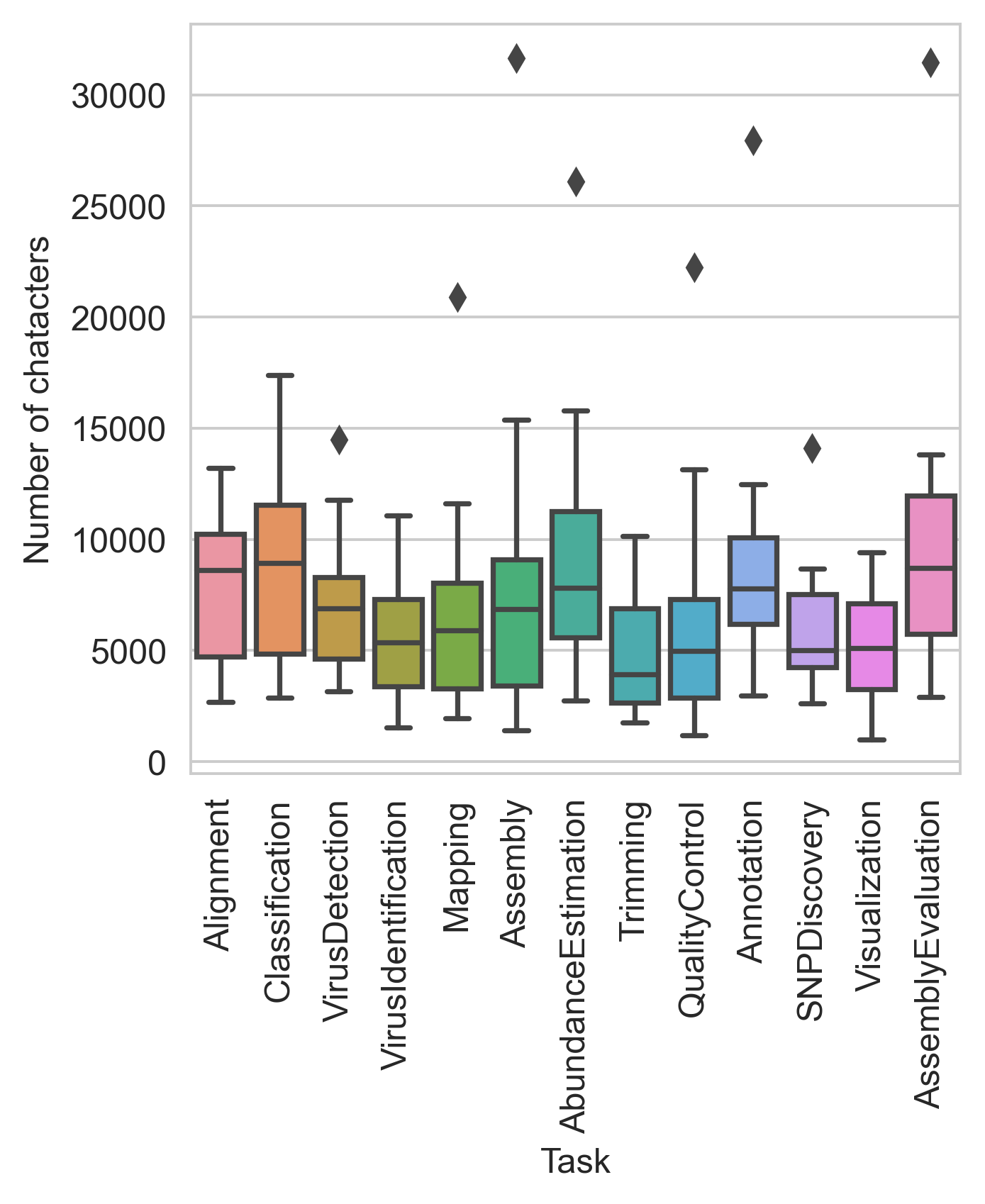}}\\
\caption{Distribution of the tool descriptions length in terms of the number of characters by task/category, (a) unprocessed text, (b) pre-processed text - ``methods only'' dataset.}\label{sfig:processedVSunprocessed_M_textsizeC_distribution_byclass}
\end{figure}

\begin{figure}[!htb]
\centering
\subfloat[]{\includegraphics[width=0.4\linewidth]{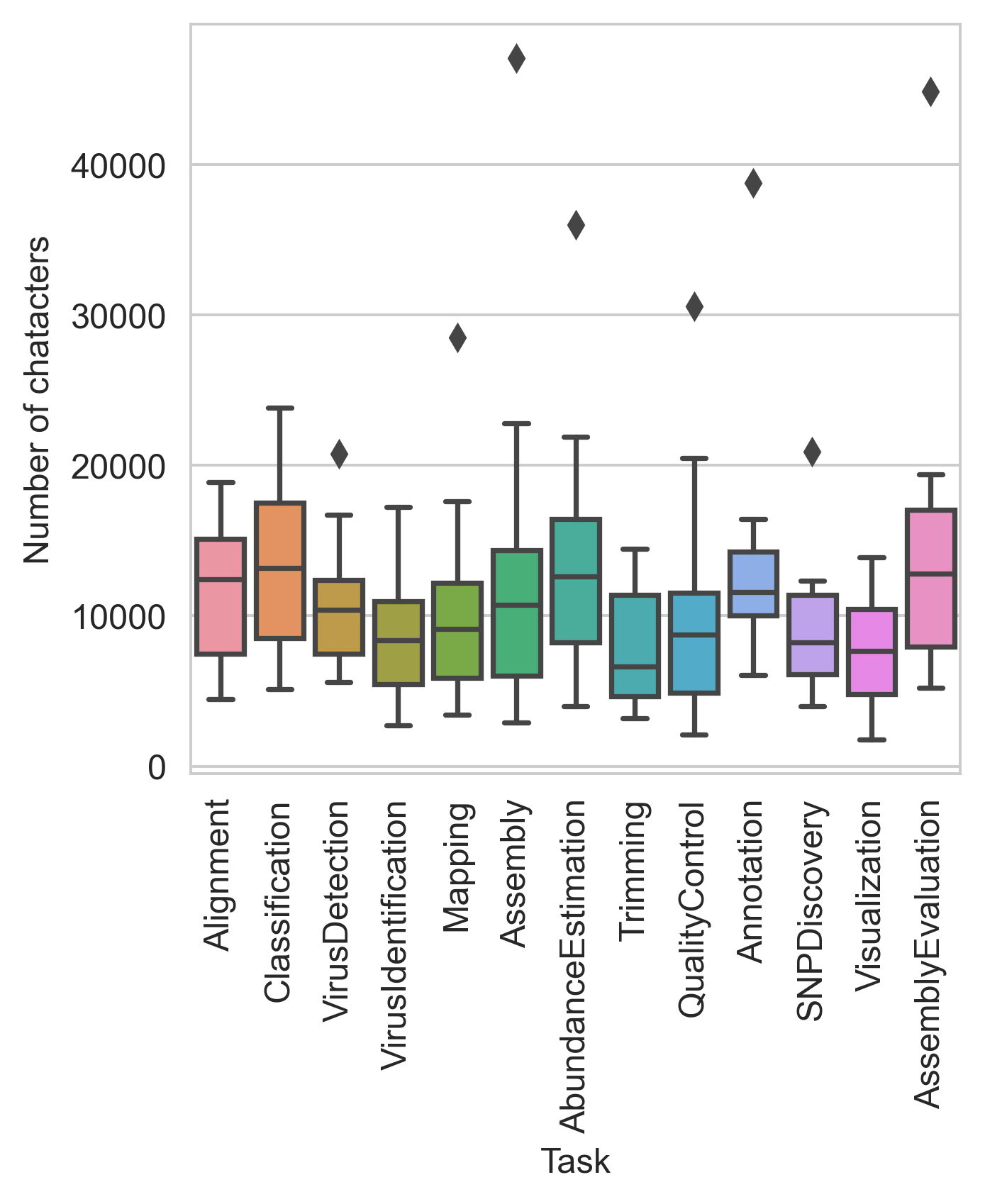}}
\subfloat[]{\includegraphics[width=0.4\linewidth]{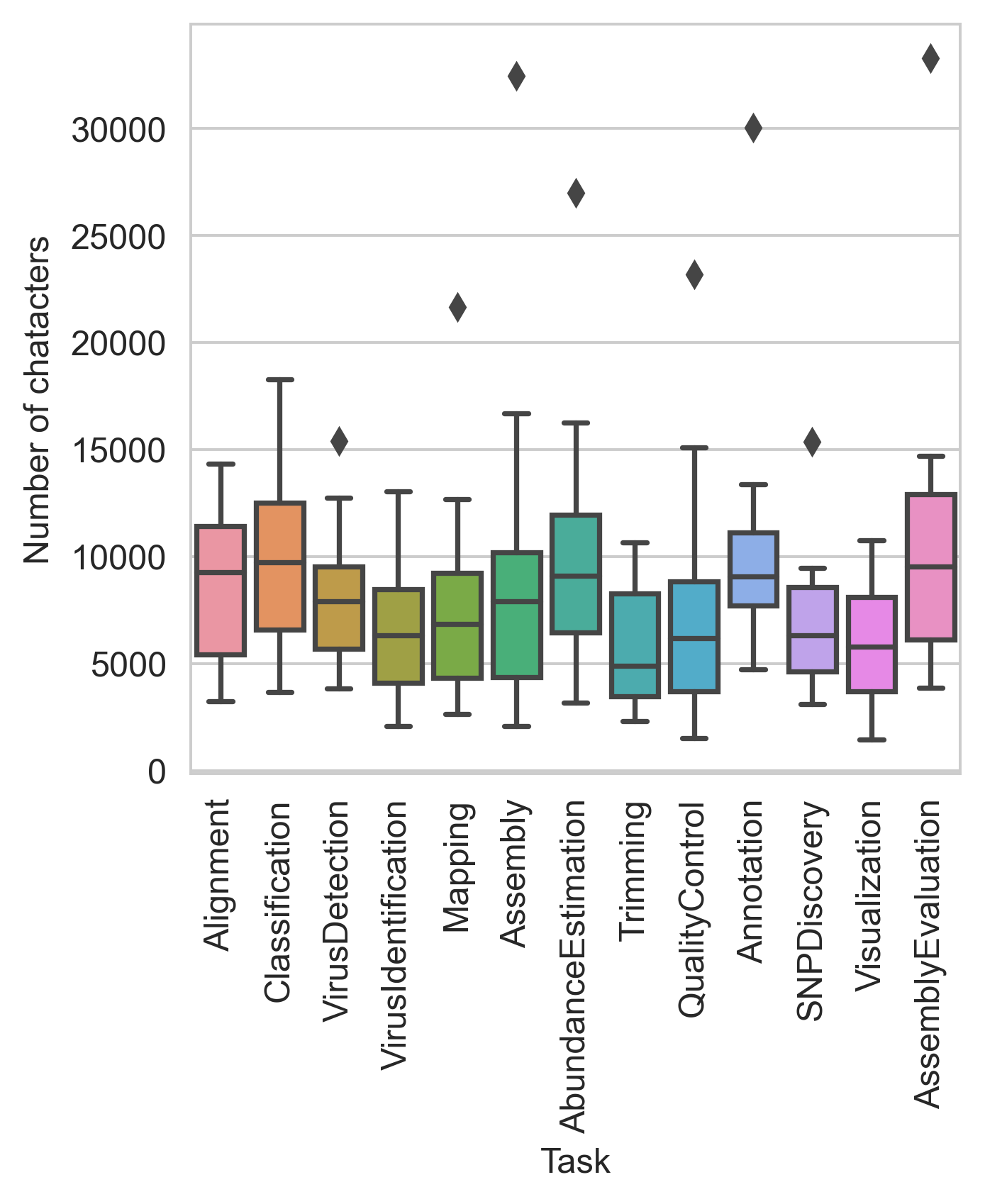}}\\
\caption{Distribution of the tool descriptions length in terms of the number of characters by task/category, (a) unprocessed text, (b) pre-processed text - ``abstracts+methods'' dataset.}\label{sfig:processedVSunprocessed_AM_textsizeC_distribution_byclass}
\end{figure}

\noindent The text processing steps reduced the number of characters in the documents, but the influence on the number of words was negligible. Text processing did not alter the distribution of the three datasets, see Supplementary Figures~\ref{sfig:processedVSunprocessed_textsize_distribution_words}, \ref{sfig:processedVSunprocessed_textsize_distribution_characters}, \ref{sfig:processedVSunprocessed_A_textsizeC_distribution_byclass}, \ref{sfig:processedVSunprocessed_M_textsizeC_distribution_byclass}, and \ref{sfig:processedVSunprocessed_AM_textsizeC_distribution_byclass}.

\subsection{Raw tool description datasets}
\subsubsection*{Supplementary data files}
Dataset S1. Raw Tool Description Abstracts Only.\\
Dataset S2. Raw Tool Description Methods Only.\\
Dataset S3. Raw Tool Description Abstracts+Methods.\\
\subsubsection*{Description}
Throughout the manual curation process of the ``methods only'' dataset, we attempted to standardize the information contained in this description as much as possible, reducing the effect of different journals' writing requirements and authors' writing styles. We searched for natural language text chunks in a given tool publication that describe the tool in great detail: techniques, algorithms, approaches, and processes. We focused on the methods section and other comparable parts of the publication. We omitted details that involved mathematical explanations and formulae from the tool descriptions. We removed any remaining mathematical symbols or equations during the text pre-processing stage.

For example, consider the description of the ``cuBLASTP'' tool from the ``methods only'' dataset. The journal publication of this tool has the following sections: (1) Introduction, (2) Background, (3) Related Work, (4) Design of Fine-Grained Blastp, (5) Performance Evaluation, and (6) Conclusion and Future Work. Therefore, we extracted the text description of this tool from the different sub-sections of the ``Design of a Fine-Grained Blastp'' section. The same approach was taken in 56 out of the 224 tool publications in our dataset, where the section explicitly titled ``Methods'' or similar (\emph{e.g.} ``Online methods'', ``Materials and methods'', ``Methods and implementation'', ``Systems and methods'', ``Methods and technologies'', ``Methodology'', \emph{etc}.) was absent.

In some publications, the ``Methods'' section is not a part of the main document; it appears in the supplementary materials. Sometimes the ``Methods'' section is present; however, it includes only technical details about the implementation of the tools (see Supplementary Figure~\ref{sfig:TechnicalMethodSection}, taken from~\cite{JosephFRyan2014}). In this case, we also collected the tool's description from other sections. In other cases, where the ``Methods'' section was present in the publication and contained an explanation of how the tool works with only a few technical details, the entire section was considered a tool description. We decided case-by-case how much technical detail was acceptable and what was excessive.

All of the datasets include the title of the tool publication. We stored the title as the first sentence of the tool description. For the ``abstracts only'' dataset, we did not perform any manual selection of text parts from the publications' abstracts as this section was consistent throughout the publications. We categorized the tools into the 13 classes of tools' tasks based on the authors' claims on the tool's primary use (11 \emph{EDAM} operations and 2 additional classes: ``Virus detection'' and ``Virus identification'').

\subsubsection*{Supplementary Figures}
\begin{figure}[H]
\subfloat[]{\includegraphics[width=0.5\linewidth]{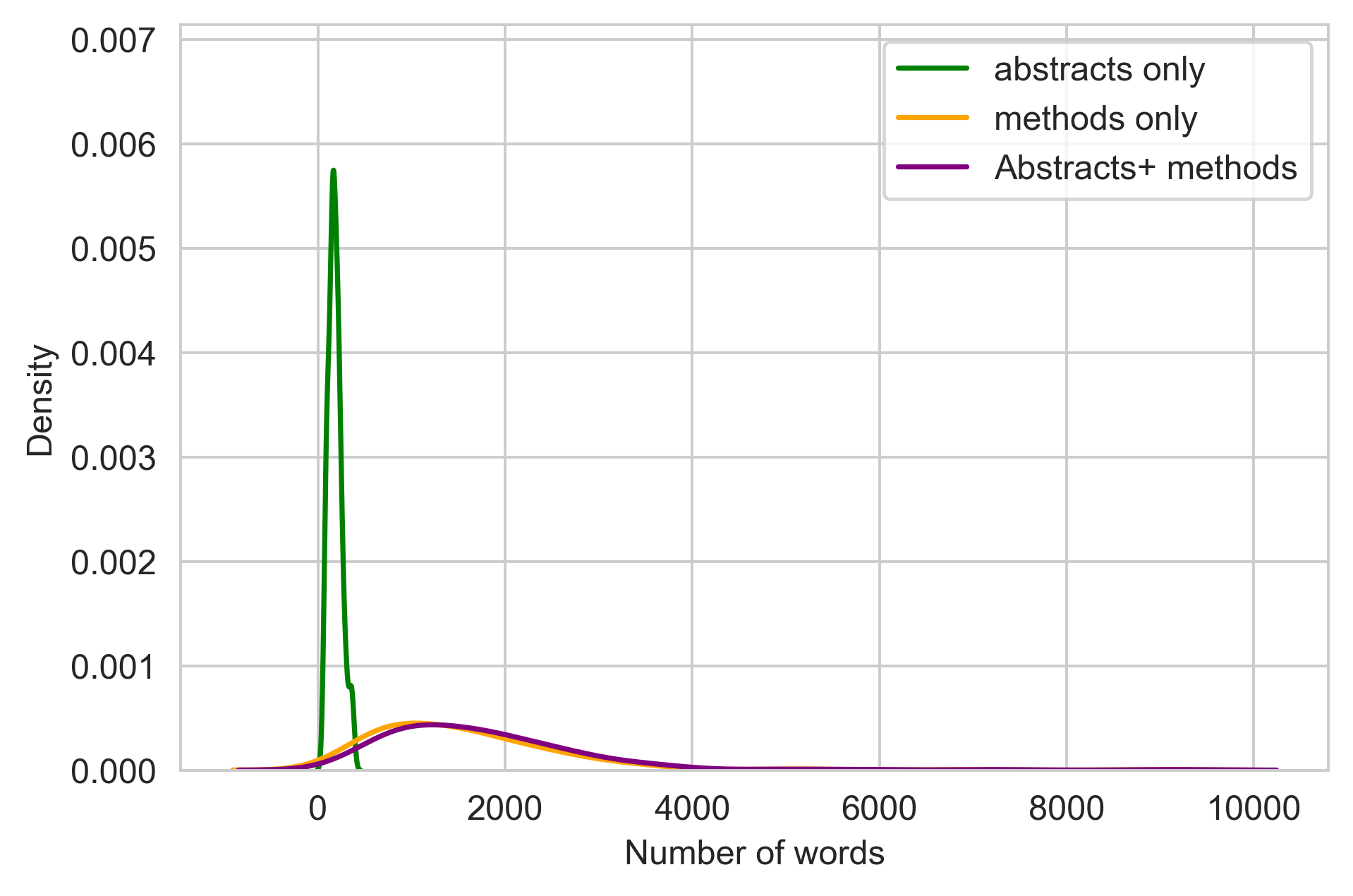}}
\subfloat[]{\includegraphics[width=0.5\linewidth]{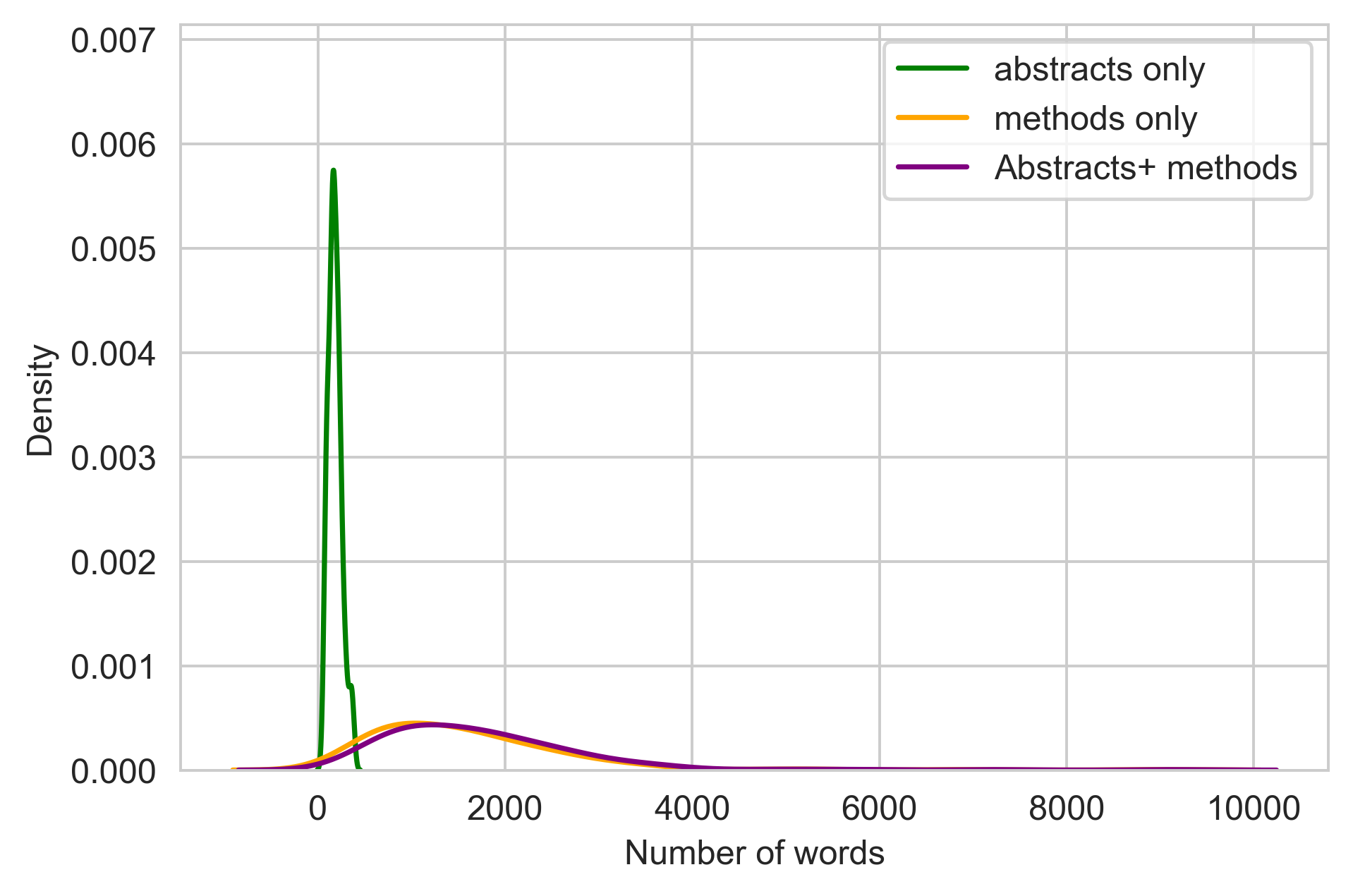}}\\
\caption{Distribution of the tool descriptions length in terms of the number of words by dataset, on the (a) unprocessed text and (b) pre-processed text.}
\label{sfig:processedVSunprocessed_textsize_distribution_words}
\end{figure}

\begin{figure}[H]
\centerline{\fbox{\includegraphics[scale=0.6]{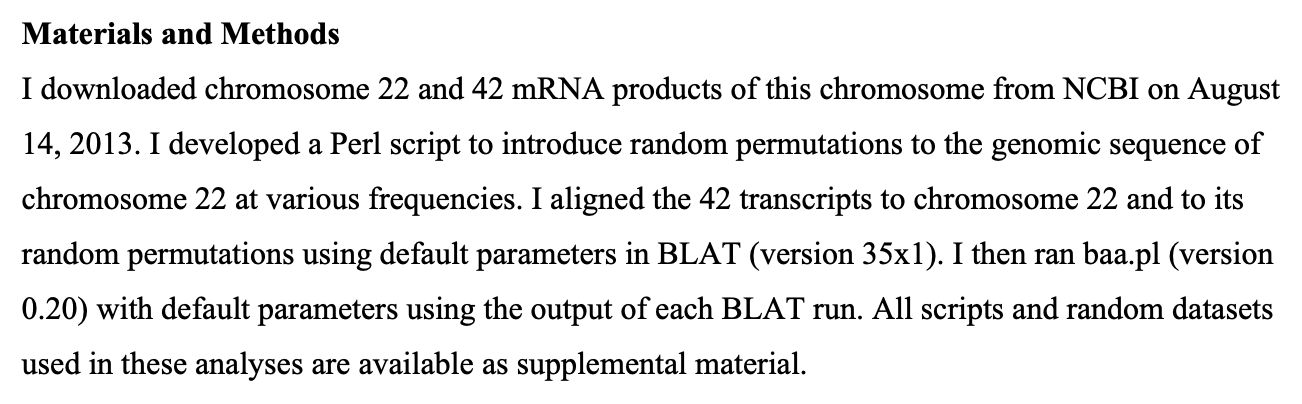}}}
\caption{Example of technical content of the ``Materials and Methods'' section, taken from~\cite{JosephFRyan2014} titled ``Baa.pl: A tool to evaluate de novo genome assemblies with RNA transcripts sets''. The tool is assigned to the ``Assembly evaluation'' task class.}\label{sfig:TechnicalMethodSection}
\end{figure}

\subsection{Pre-processed tool description datasets}
\subsubsection*{Supplementary data files}
Dataset S4. Pre-processed Tool Description Abstracts Only.\\
Dataset S5. Pre-processed Tool Description Methods Only.\\
Dataset S6. Pre-processed Tool Description Abstracts+Methods.\\
\subsubsection*{Description}
Pre-processed datasets contain raw tool descriptions pre-processed in the following way:
\begin{itemize}\item down-casing,
    \item removing special characters,
    \item punctuation signs,
    \item possessive pronouns,
    \item numbers,
    \item links,
    \item words composed of one or two letters,
    \item words starting with digits,
    \item stop words removal,
    \item and lemmatization.
\end{itemize}

\noindent We assigned a numerical ID to each of the 13 classes.

\subsubsection*{Supplementary Figures}

\clearpage

\section{Methods} \label{s:methods}
In this section, we explain the methods used to build the models, including the hyperparameters values tested during cross-validation and the different pooling strategies evaluated on BERT. We also list the text embedding datasets generated by the different embedding strategies. 
\subsection{Model hyperparameters}
\subsubsection*{Description}
During the repeated nested cross-validation process, we optimized the parameters of the classifiers (Logistic Regression, Random Forest, Naive Bayes) on the datasets generated by each embedding method. The embedding methods were not fine-tuned, except for TF-IDF. We report on the scikit-learn’s \emph{GridSearchCV}~\cite{scikit-learn} hyperparameters names and corresponding settings tested in Supplementary Table~\ref{stab:CrossValidationHyperparameters}. 

\subsubsection*{Tables}
\begin{table}[H]
\caption{Hyperparameter settings tested during the cross-validation step for the TF-IDF embedding and the classification algorithms. }\label{stab:CrossValidationHyperparameters}
\scriptsize
\begin{tabular}{llll}\toprule
Algorithm & Hyperparameter & Value set \\\midrule
\multirow{5}{*}{TF-IDF} & max\_features & [40, 140, 180, 220] \\
& ngram\_range & [(1, 1), (1, 2), (2, 2)] \\
& min\_df & [0.001, 0.01, 0.1] \\
& max\_df & [0.5, 0.6, 0.7, 0.8, 0.9] \\
& stop\_words & [None, 'english'] \\
\multirow{3}{*}{Logistic Regression} & C & [0.1, 0.2, 0.3, 0.4, 0.5, 0.6, 0.7, 0.8, 0.9, 1. ] \\
& solver & ['newton-cg', 'sag', 'saga', 'lbfgs'] \\
& class\_weight & ['balanced', None] \\
\multirow{6}{*}{Random Forest} & max\_depth & [100] \\
& min\_samples\_split & [2, 5, 10] \\
& min\_samples\_leaf & [1, 2, 4] \\
& max\_leaf\_nodes & [None] \\
& max\_features & ['auto', 'sqrt'] \\
& bootstrap & [True, False] \\
Naive Bayes & var\_smoothing & [0.00000001, 0.000000001, 0.00000001] \\
\bottomrule
\end{tabular}
\end{table}

\subsection{BERT pooling strategies}
\subsubsection*{Description}
In this study, we used the feature-based approach with all the embedding methods we tested to extract fixed features based on pre-trained methods. In addition to the different BERT-based models, which use different training corpora, architectures, and training methodologies, we also tested the ``best layer to use'' of the BERT model by extracting the activations from different layers. We deduced that the sum (BERTSL4) and concatenation (BERTCL4) of the last four layers work better than the last layer (BERT-st). However, the sum of the second to last layer (BERTS2L) is the best pooling strategy. For details, see Supplementary Figure~\ref{sfig:abstractsPRCscores}. Although these results are specific to the problem of classifying bioinformatics text based on the abstract descriptions of tools that we address in our study, they are still similar to the results of the original BERT paper~\cite{devlin2019bert}. Some specific choices of layers outperform the last layer, which can be explained by the bias introduced by the proximity of the last layer to the target functions, \emph{e.g.}, next sentence prediction pre-training task. 
Due to the high noise in the text data from the ``methods only'' dataset, we did not consider its results.

\subsection{Text embedding datasets}
\subsubsection*{Supplementary data files}
Dataset S7. BioBERT Hugging Face Abstracts Only.\\
Dataset S8. BERTSL4 Abstracts Only.\\
Dataset S9. BERTCL4 Abstracts Only.\\
Dataset S10. BERTS2L Abstracts Only.\\
Dataset S11. BERT Sentence Transformer Abstracts Only.\\
Dataset S12. BioBERT NLU Abstracts Only.\\
Dataset S13. ELECTRA Hugging Face Abstracts Only.\\
Dataset S14. ELECTRAmed Hugging Face Abstracts Only.\\
Dataset S15. RoBERTa Sentence Transformer AbstractsOnly.\\
Dataset S16. ELMO AbstractsOnly.\\
Dataset S17. XLNET NLU Abstracts Only.\\
Dataset S18. GLOVE NLU Abstracts Only.\\
Dataset S19. ELECTRAmed Sentence Transformer Abstracts Only.\\
Dataset S20. ELECTRA NLU Abstracts Only.\\
Dataset S21. BERTSL4 Methods Only.\\
Dataset S22. BERTCL4 Methods Only.\\
Dataset S23. BERTS2L Methods Only.\\
Dataset S24. BERT Sentence Transformer Methods Only.\\
Dataset S25. BioBERT NLU Methods Only.\\
Dataset S26. RoBERTa Sentence Transformer Methods Only.\\
Dataset S27. ELMO Methods Only.\\
Dataset S28. XLNET NLU Methods Only.\\
Dataset S29. GLOVE NLU Methods Only.\\
Dataset S30. ELECTRAmed Sentence Transformer Methods Only.\\
Dataset S31. ELECTRA NLU Methods Only.\\
Dataset S32. BERTSL4 Abstracts+Methods.\\
Dataset S33. BERTCL4 Abstracts+Methods.\\
Dataset S34. BERTS2L Abstracts+Methods.\\
Dataset S35. BERT Sentence Transformer Abstracts+Methods.\\
Dataset S36. BioBERT NLU Abstracts+Methods.\\
Dataset S37. RoBERTa Sentence Transformer Abstracts+Methods.\\
Dataset S38. ELMO Abstracts+Methods.\\
Dataset S39. XLNET NLU Abstracts+Methods.\\
Dataset S40. GLOVE NLU Abstracts+Methods.\\
Dataset S41. ELECTRAmed Sentence Transformer Abstracts+Methods.\\
Dataset S42. ELECTRA NLU Abstracts+Methods.\\
\subsubsection*{Description}
This supplement section contains text vector representations generated by different Embedding strategies for the three pre-processed tool description datasets: ``abstracts only'', ``methods only'' and ``abstracts+methods''.

\subsubsection*{Tables}
\begin{table}[H]
\caption{Overview of the vector size generated by each embedding technique. The size of the TF-IDF vector varies according to the hyperparameter tuning results.}\label{stab:EmbeddingVectorSize}
\scriptsize
{\begin{tabular}{@{}ll@{}}\toprule 
Embedding Method & Embedding size \\\midrule
TF-IDF & - \\
GLOVE & 100 \\
ELMO & 1024 \\
BERT & 768 \\
BERT S2L & 768 \\
BERT SL4 & 768 \\
BERT CL4 & 3072 \\
BioBERT & 768 \\
RoBERTa & 768\\
XLNET & 768 \\
ELECTRA & 256 \\ \bottomrule
\end{tabular}}{}
\end{table}

\section{Results}
In this section, we report on the models' best hyperparameters setting, the performance metrics of all models, including the baseline model, the classification probabilities produced by the best models, and a detailed table of the best model misclassifications. We also examined the results of the top three models and the worst model.

\subsection{Baseline model - Majority classifier} \label{s:baseline}
\subsubsection*{Description}
\begin{figure}[H]
\centerline{\includegraphics[scale=0.6]{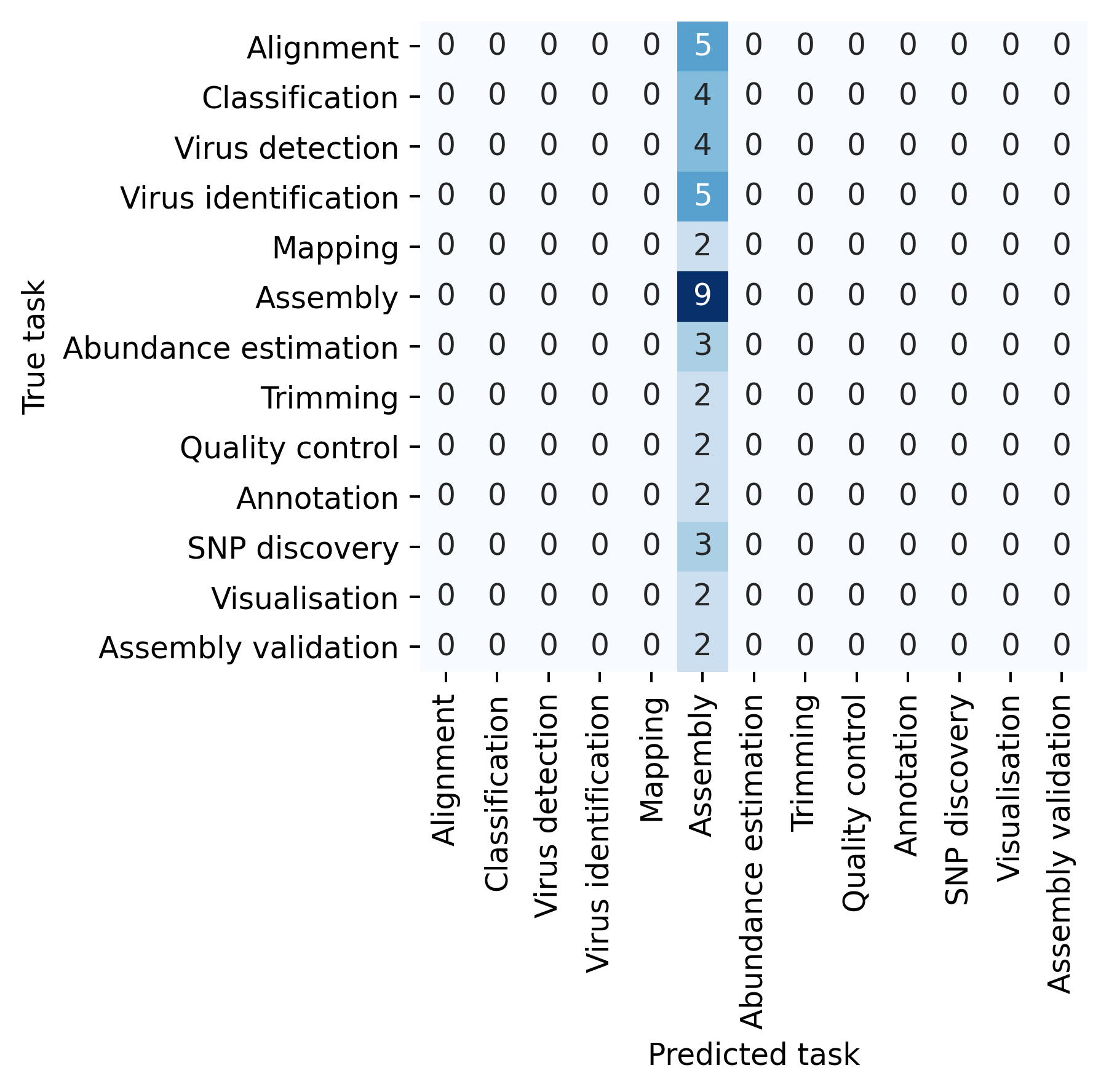}}
\caption{Majority classifier test set confusion matrix - similar for all datasets.}\label{sfig:majorityClassifierCM}
\end{figure}

The baseline model classifies all the tools as Assembly tools, the majority class, which represents 19.6\% of the examples in our datasets, see Supplementary Table~\ref{stab:MajorityClassifierMetrics} and Figure~\ref{sfig:majorityClassifierCM}.

\subsubsection*{Tables}
\begin{table}[H]
\caption{\label{stab:MajorityClassifierMetrics}Majority Classifier metrics - independent of dataset and embedding method.}
\scriptsize
{\begin{tabular}{@{}ll@{}}\toprule 
Evaluation metric & Score \\\midrule
Training set Accuracy & 0.19 \\
Test set Accuracy & 0.20 \\
AUC-ROC & 0.50 \\
AUC-PR & 0.10 \\
Precision & 0.04 \\
Recall & 0.20 \\
Fscore & 0.06 \\ \bottomrule
\end{tabular}}{}
\end{table}

\subsubsection*{Supplementary Figures}

\subsection{Model performance}
\label{subs:model-performance}
\subsubsection*{Supplementary data files}
Data Table S43. Performance of tested models.
\subsubsection*{Description}
For each dataset (``abstracts only'', ``methods only'', and ``abstracts+methods''), we report the following scores of all trained models on the independent test set, ordered by the Area Under the Receiver Operating Characteristic Curve (\emph{AUC-ROC}) score:
\begin{itemize}
    \item training set \emph{Accuracy}, 
    \item independent test set \emph{Accuracy}, 
    \item (\emph{AUC-ROC}) score, 
    \item \emph{Precision}, 
    \item \emph{Recall}, and
    \item \emph{F1-score}. 
\end{itemize}
The results are discussed in Supplementary Section~\ref{subs:model-comparison}.

\subsection{Models' best hyperparameters}
\subsubsection*{Supplementary data files}
Data Table S44. Hyperparameters set.\\
Data Table S45. Model best hyperparameters.
\subsubsection*{Description}
For each dataset (``abstracts only'', ``methods only'', and ``abstracts+methods''), we report the set of tested values for each hyperparameter as well as the best hyperparameter values of the three classifiers (Logistic Regression, Random Forest, and Naive Bayes) when trained on different text vector representations.

\subsection{Best models' class probabilities}
\subsubsection*{Supplementary data files}
Data Table S46. Best models predicted class probabilities.
\subsubsection*{Description}
In this supplementary data, we report the classification probabilities of each class on the independent test set composed of 45 examples for the models that achieved the best (\emph{AUC-PR}). The datasets are as follows:
\begin{enumerate}
\item Logistic Regression on BioBERT-hf embeddings of ``abstracts only'' dataset.
\item Logistic Regression on BERTS2L embeddings of ``abstracts+methods'' dataset.
\item Logistic Regression on BERTS2L embeddings of ``abstracts only'' dataset, and Logistic Regression on BERTSL4 of ``abstracts+methods'' dataset.
\end{enumerate}

\subsection{Model comparison}
\label{subs:model-comparison}
\subsubsection*{Description}
All trained models were evaluated as described in Supplementary Subsection~\ref{subs:model-performance}, see Supplementary Figures~\ref{sfig:LR+BERTS2L_abstmethd}, \ref{sfig:LR+BERTS2L_abstonly}, \ref{sfig:LR+BERTSL4_abstmethd}, and \ref{sfig:LR+ELECTRA-NLU_methonly}. We determined that Logistic Regression on NLU library implementation of ELECTRA embeddings of the ``methods only'' dataset performed the worst with an (\emph{AUC-PR}) score of 0.13. The model missed 8 out of 13 classes: ``(Sequence) alignment'', ``(Taxonomic) classification'', ``Virus identification'', ``Mapping'', ``(Sequence) trimming'',  ``SNP-Discovery'', ``(Sequence) annotation'' and ``(Sequence) assembly validation''; thus, it partially learned five categories only.
Logistic Regression on BERTS2L embeddings of the ``abstracts+methods'' dataset was the second-best model with (\emph{AUC-PR}) score of 0.84. It completely missed the instances of the ``(Sequence) annotation'', ``(Sequence) assembly'', and ``Visualization'' classes. The model also had problems distinguishing between instances from ``Virus detection'' and ``Virus identification'' classes. Oppositely, the model achieved a 100\% accuracy in classifying instances of four classes: ``Mapping'', ``(RNA-seq quantification for) abundance estimation'', ``(Sequencing) quality control'' and ``(Sequence) assembly validation'' classes.
Logistic Regression on BERTS2L embeddings of the ``abstracts only'' dataset, and the Logistic Regression on BERTSL4 embeddings of the ``abstracts+methods'' dataset,  third-best models with a 0.83 (\emph{AUC-PR}) score both, achieved similar results to the second-best model.

In addition, models were compared using the Critical Distance diagram method proposed by Demšar~\cite{demsar2006statistical}. A comparison of the best three models and the worst model against our baseline is shown in Supplementary Figure~\ref{sfig:PRCCurves}.

\begin{figure}[H]
\centerline{\includegraphics[scale=0.55]{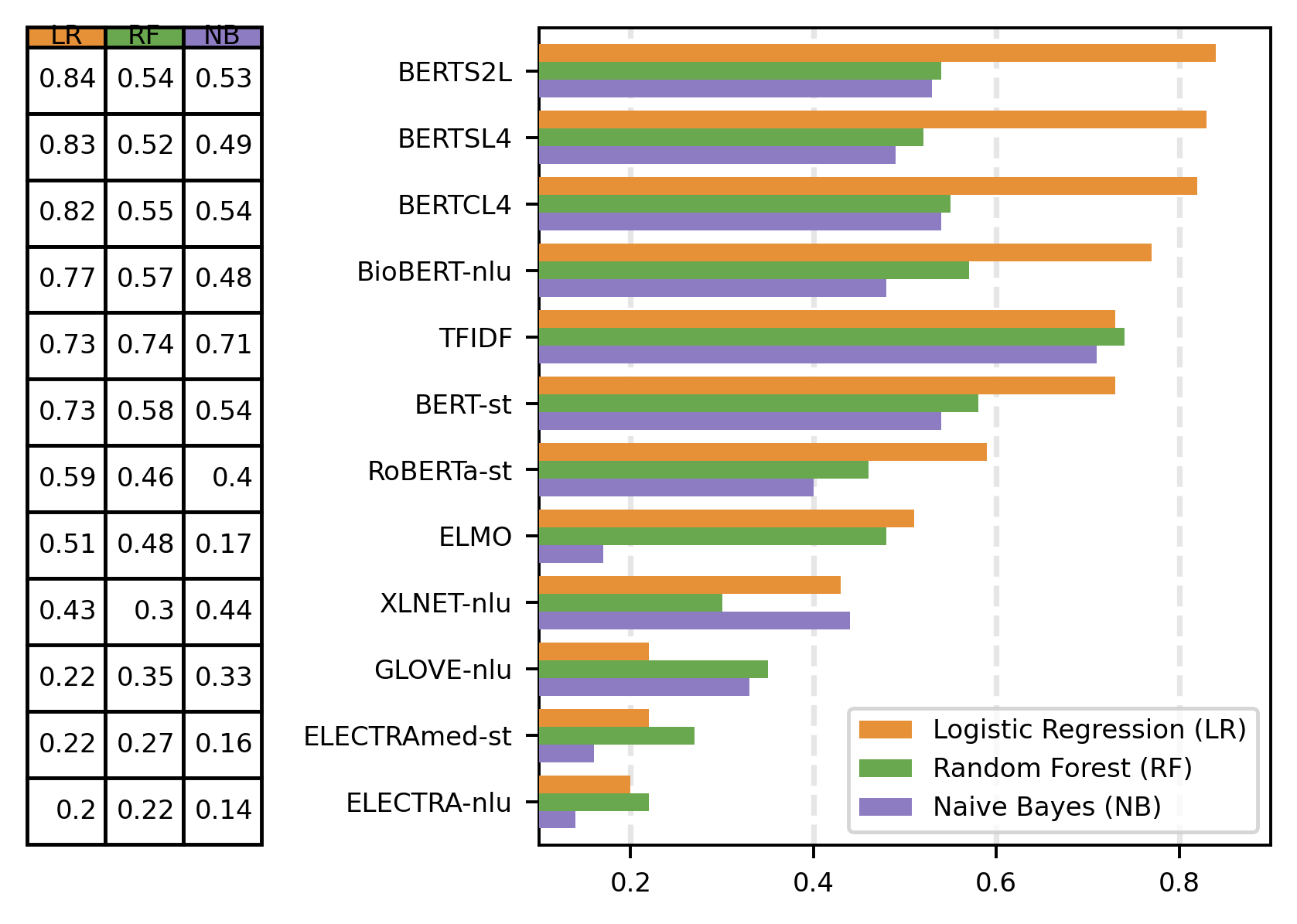}}
\caption{Classifiers' \emph{AUC-PR} scores grouped by embedding method on ``abstracts+methods'' dataset.}
\label{sfig:abstractsmethodsPRCscores}
\end{figure}

For an easier comparison of the performance of the various methods and data sources, we present the results from ``Data Table S43. Performance of tested models'' as graphs in Supplementary Figures~\ref{sfig:abstractsPRCscores}, \ref{sfig:methodsPRCscores} and~\ref{sfig:abstractsmethodsPRCscores}.

\subsubsection*{Supplementary Figures}

\begin{figure}[H]
\centering{
\subfloat[]{\includegraphics[scale=0.6]{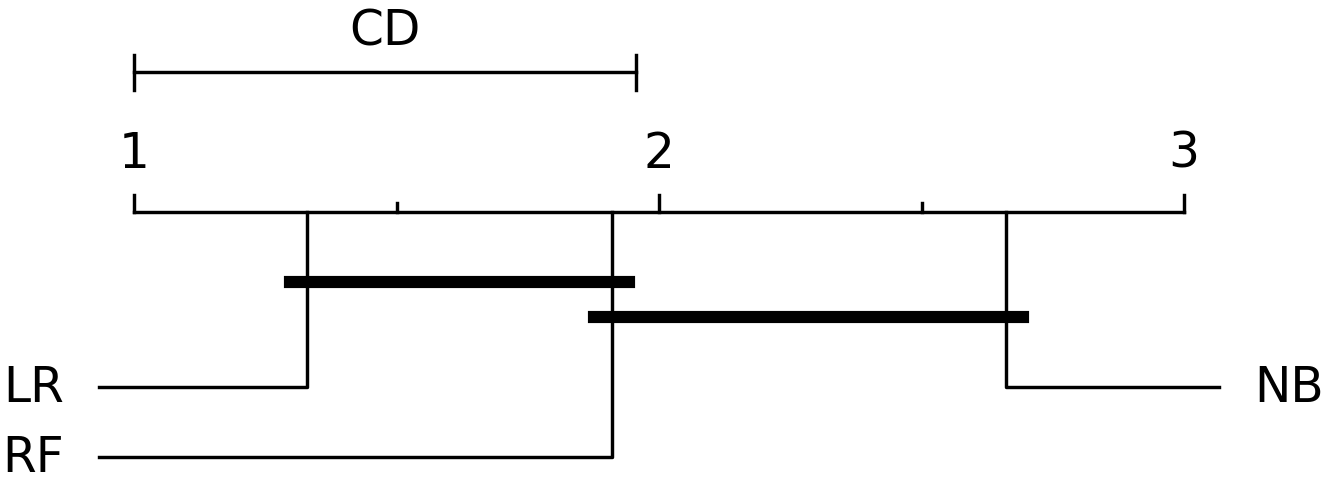}}

\subfloat[]{\includegraphics[scale=0.6]{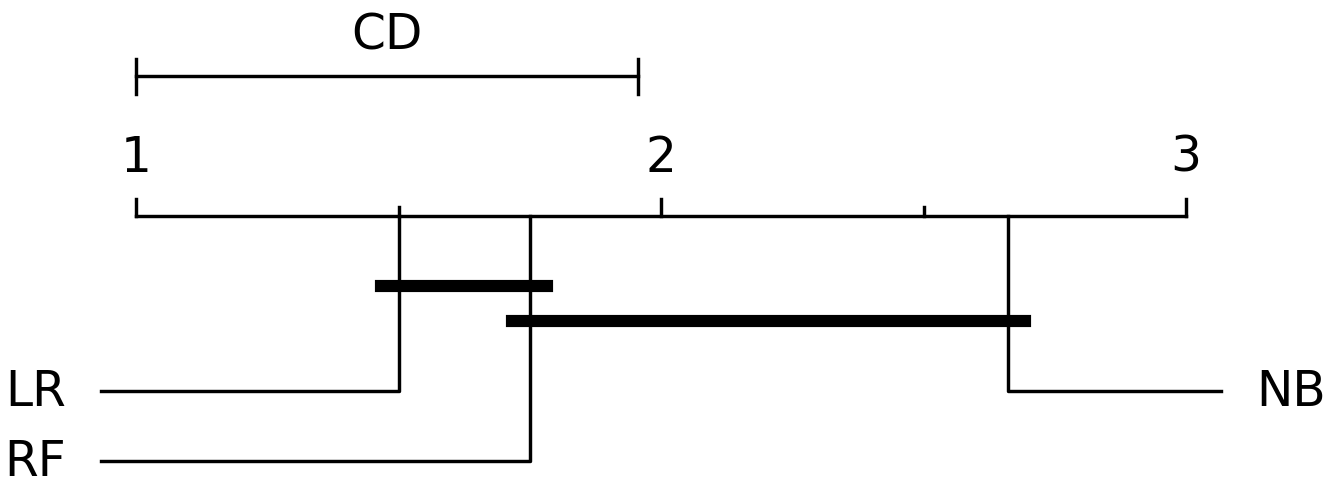}}

\subfloat[]{\includegraphics[scale=0.6]{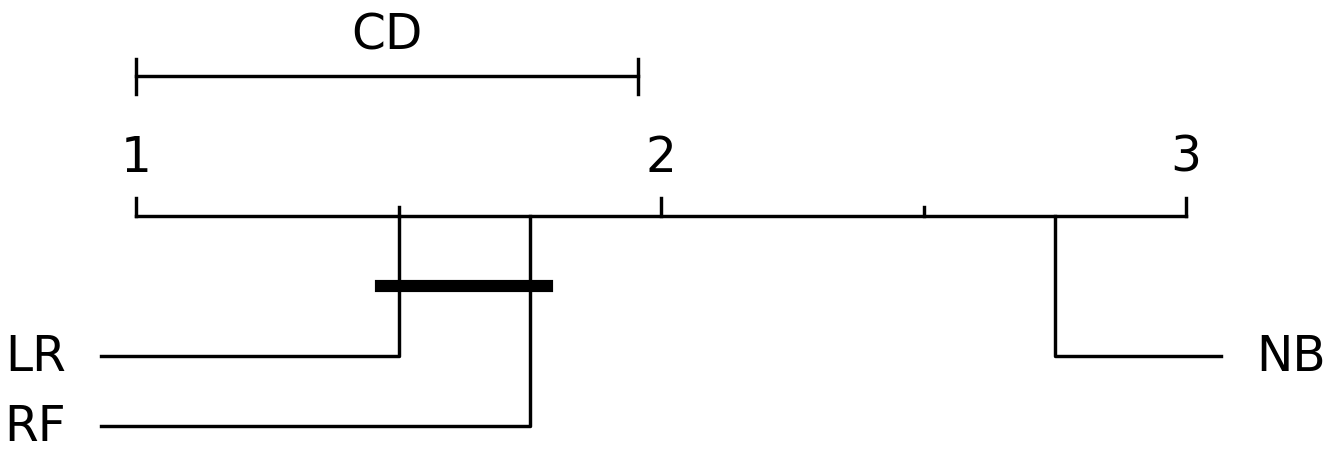}}}
\caption{Critical distance (CD) diagram for LR, RF,an NB classifiers tested on 12 embeddings. Groups of classifiers that are not significantly different are connected on the (a) ``abstracts only'', (b) ``methods only'', and (c) ``abstracts+methods'' dataset.}\label{sfig:nemenyi-abst-methd-combined}
\end{figure}

\begin{figure}[H]
\centerline{\includegraphics[scale=0.6]{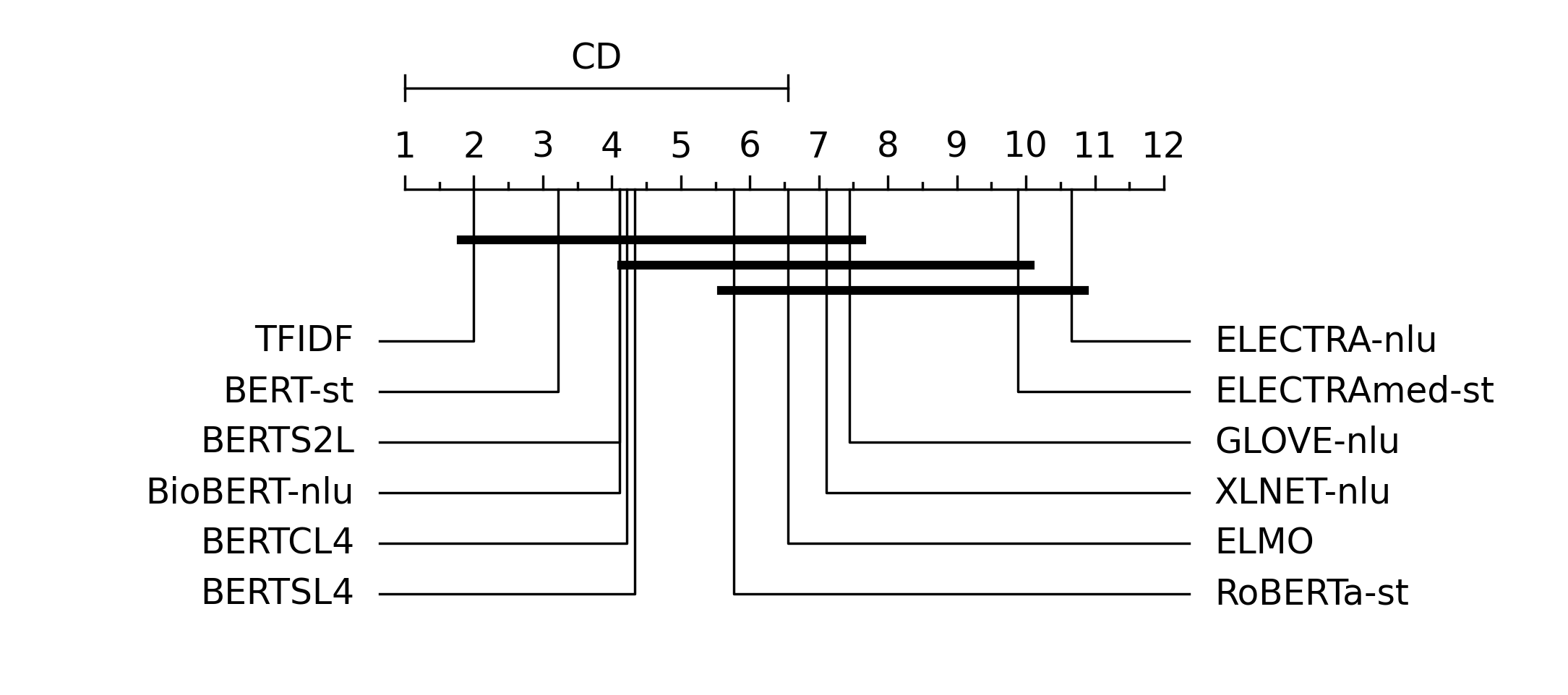}}
\caption{Comparison of embedding methods, ranked by (\emph{AUC-PR}), using the Nemenyi test. Tested on three classifiers and three datasets. Groups of embeddings that are not significantly different are connected.}\label{sfig:nemenyi-embeddings-3cx3d}
\end{figure}

\begin{figure}[H]
\centerline{\includegraphics[scale=0.42]{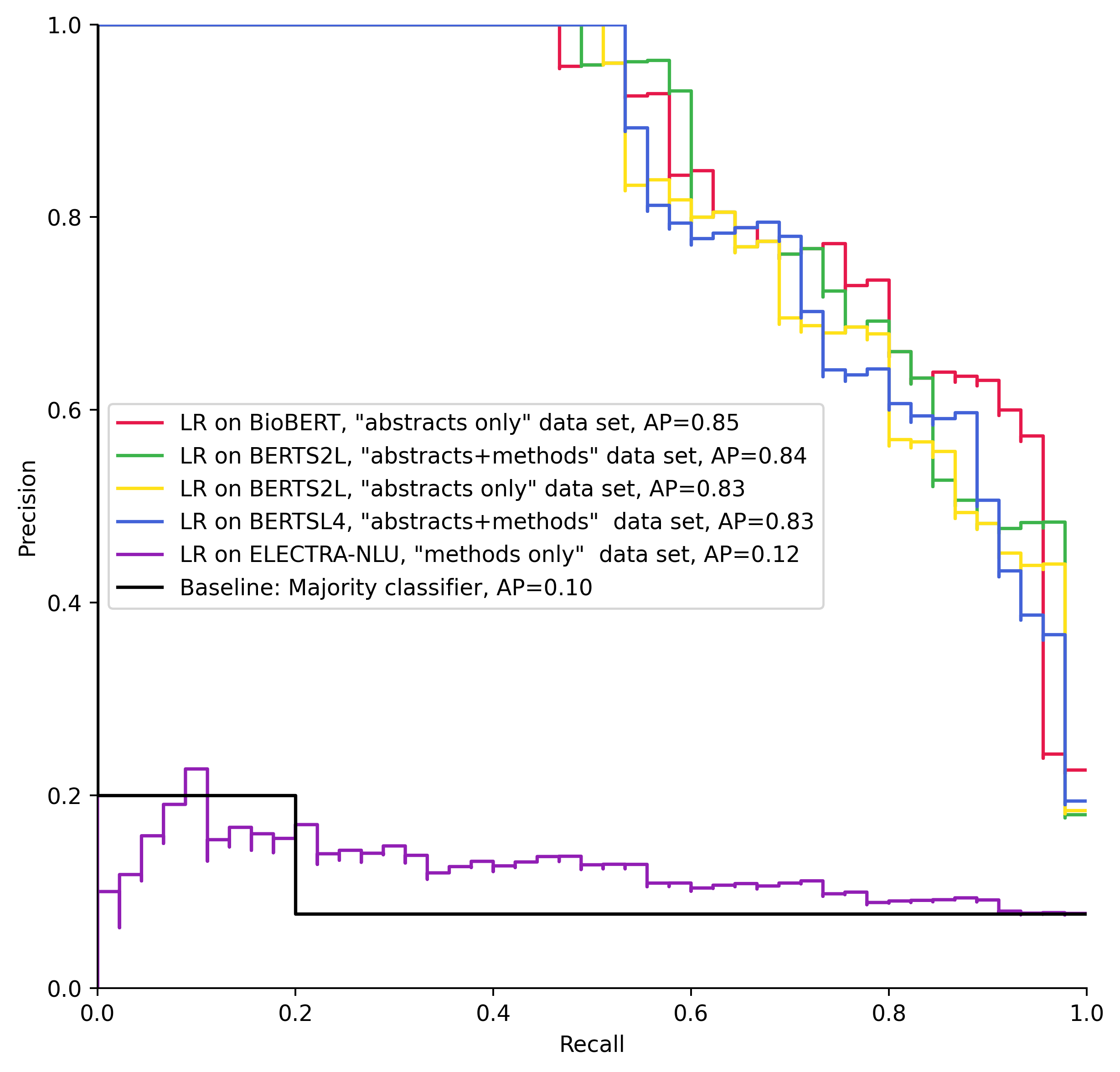}}
\caption{Average precision scores, micro-averaged over all classes. The majority classifier, the worst model (LR ELECTRA-NLU) and the top four models (third and fourth have the same performance) are shown.}\label{sfig:PRCCurves}
\end{figure}

\begin{figure}[H]
\centerline{\includegraphics[scale=0.55]{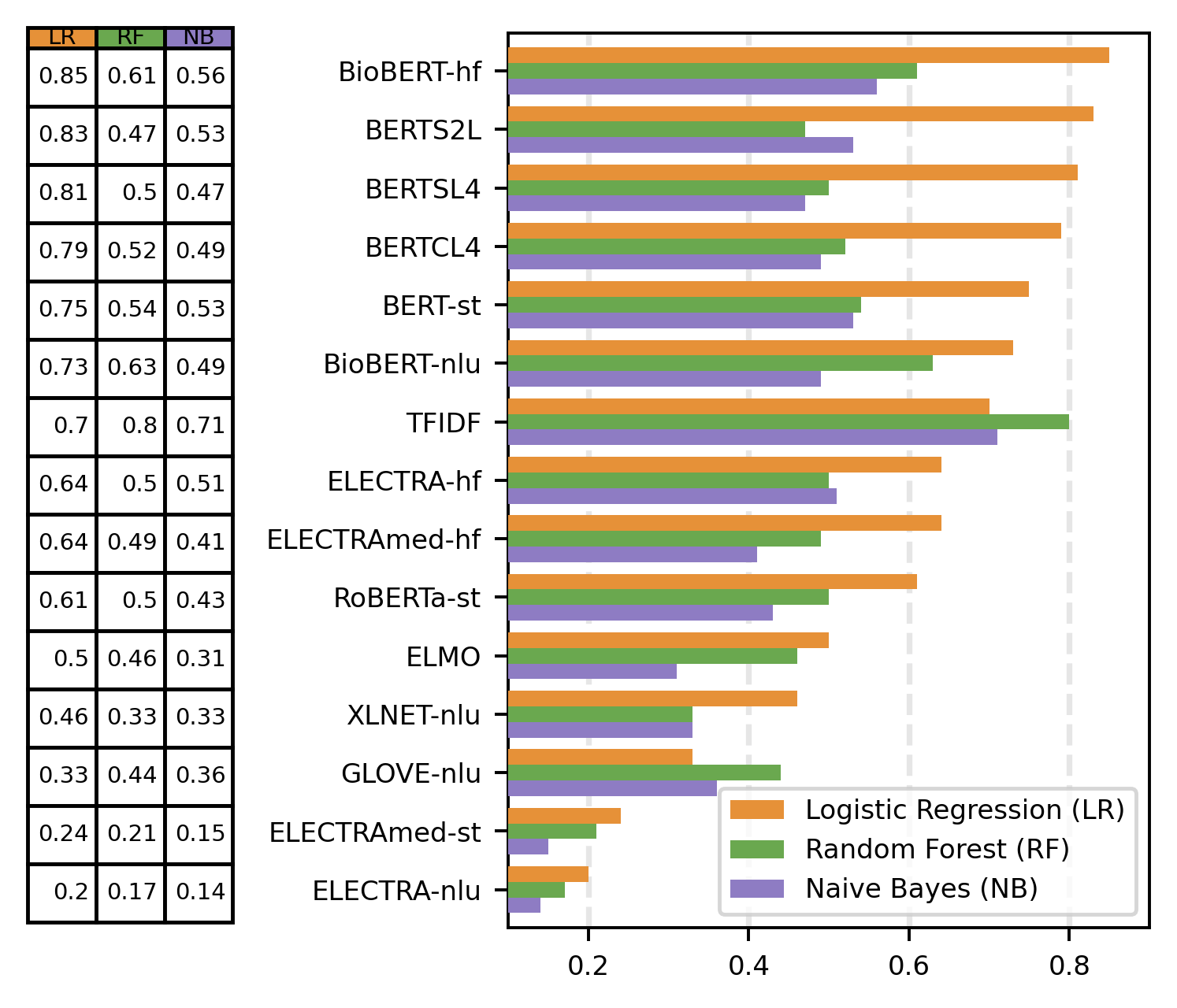}}
\caption{Classifiers' \emph{AUC-PR} scores grouped by embedding method on ``abstracts only'' dataset.}\label{sfig:abstractsPRCscores}
\end{figure}

\begin{figure}[H]
\centerline{\includegraphics[scale=0.55]{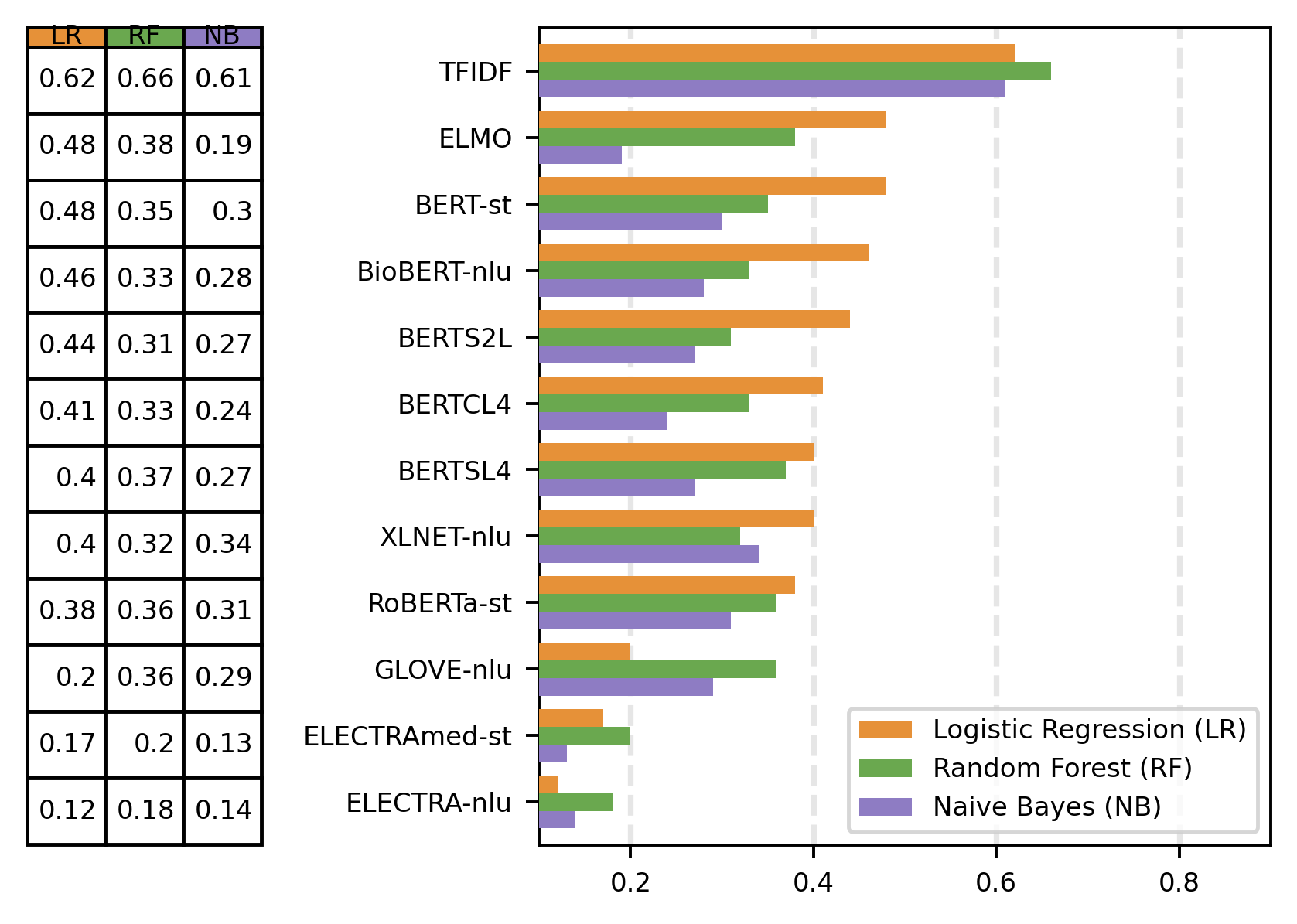}}
\caption{Classifiers' \emph{AUC-PR} scores grouped by embedding method on ``methods only'' dataset.}\label{sfig:methodsPRCscores}
\end{figure}

\begin{figure}[H]
    \centering
    \begin{minipage}{0.45\textwidth}
        \centering
        \includegraphics[scale=0.5]{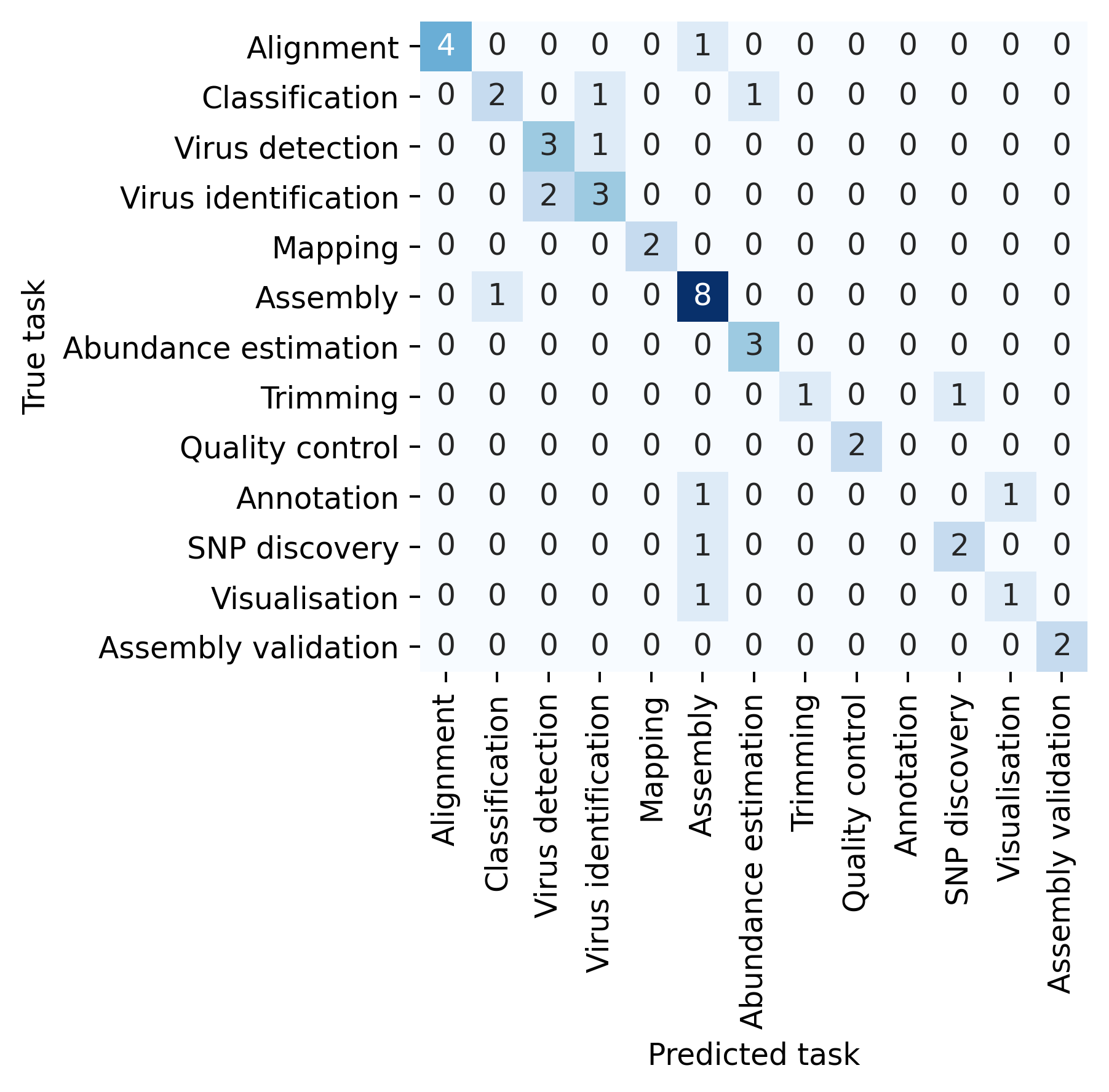}
        \caption{Logistic Regression on BERTS2L - test set confusion matrix - ``abstracts+methods'' dataset.}\label{sfig:LR+BERTS2L_abstmethd}
    \end{minipage}\hfill
    \begin{minipage}{0.45\textwidth}
        \centering
        \includegraphics[scale=0.5]{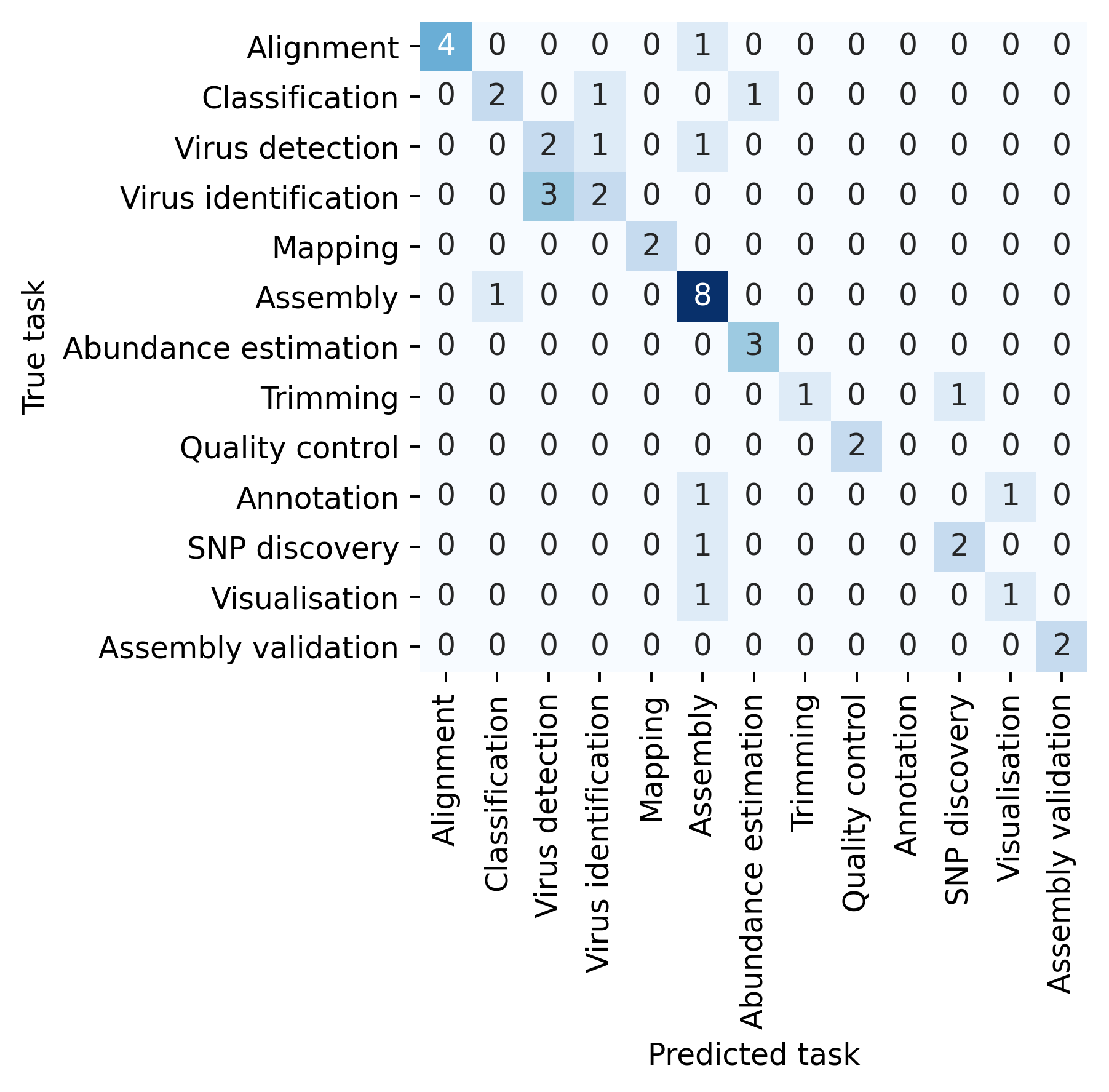}
        \caption{Logistic Regression on BERTS2L - test set confusion matrix - ``abstracts only'' dataset.}\label{sfig:LR+BERTS2L_abstonly}
    \end{minipage}
    \vspace{1cm}

    \begin{minipage}{0.45\textwidth}
        \centering
        \includegraphics[scale=0.5]{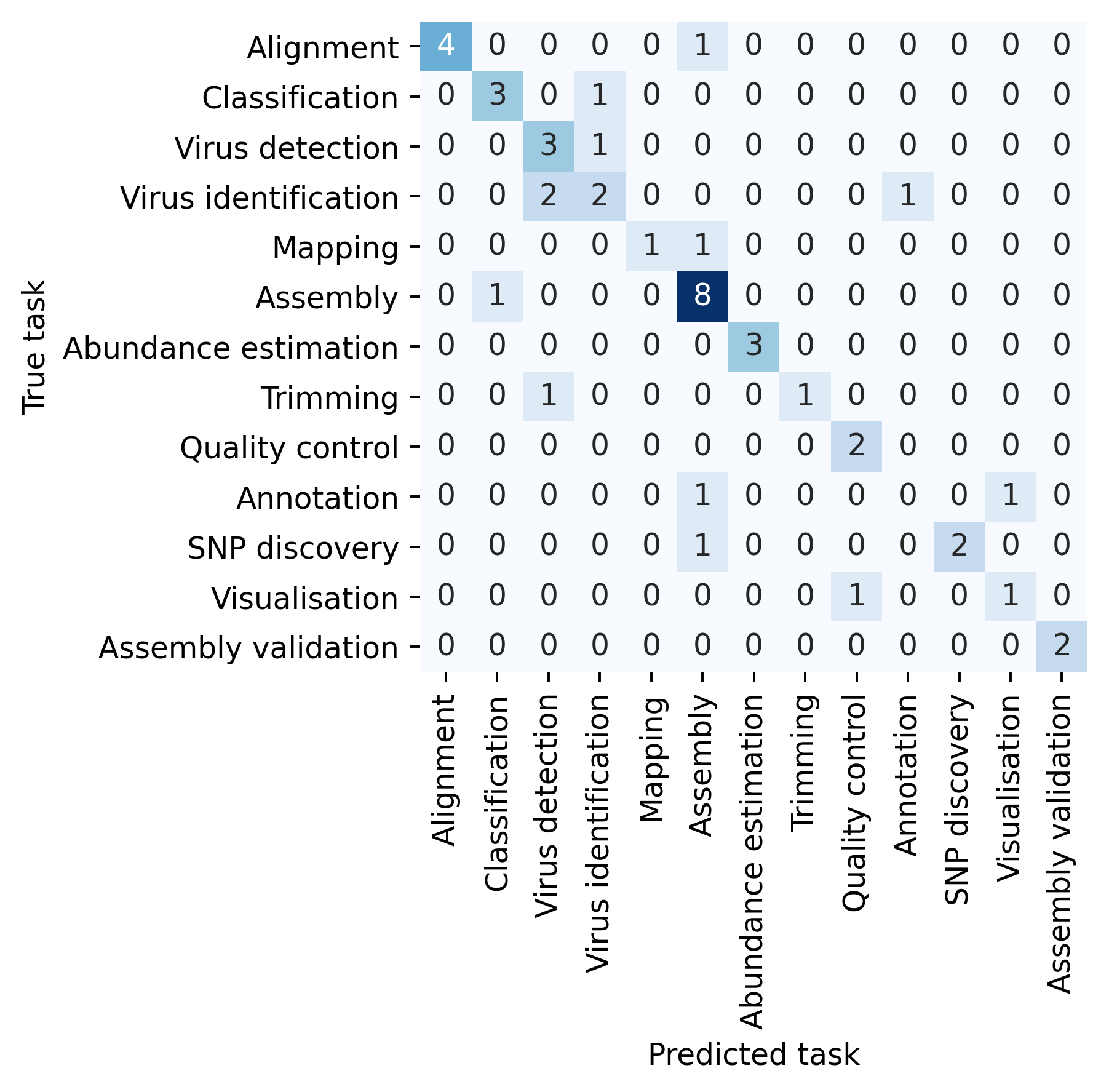}
        \caption{Logistic Regression on BERTSL4 - test set confusion matrix - ``abstracts+methods'' dataset.}\label{sfig:LR+BERTSL4_abstmethd}
    \end{minipage}\hfill
    \begin{minipage}{0.45\textwidth}
        \centering
        \includegraphics[scale=0.5]{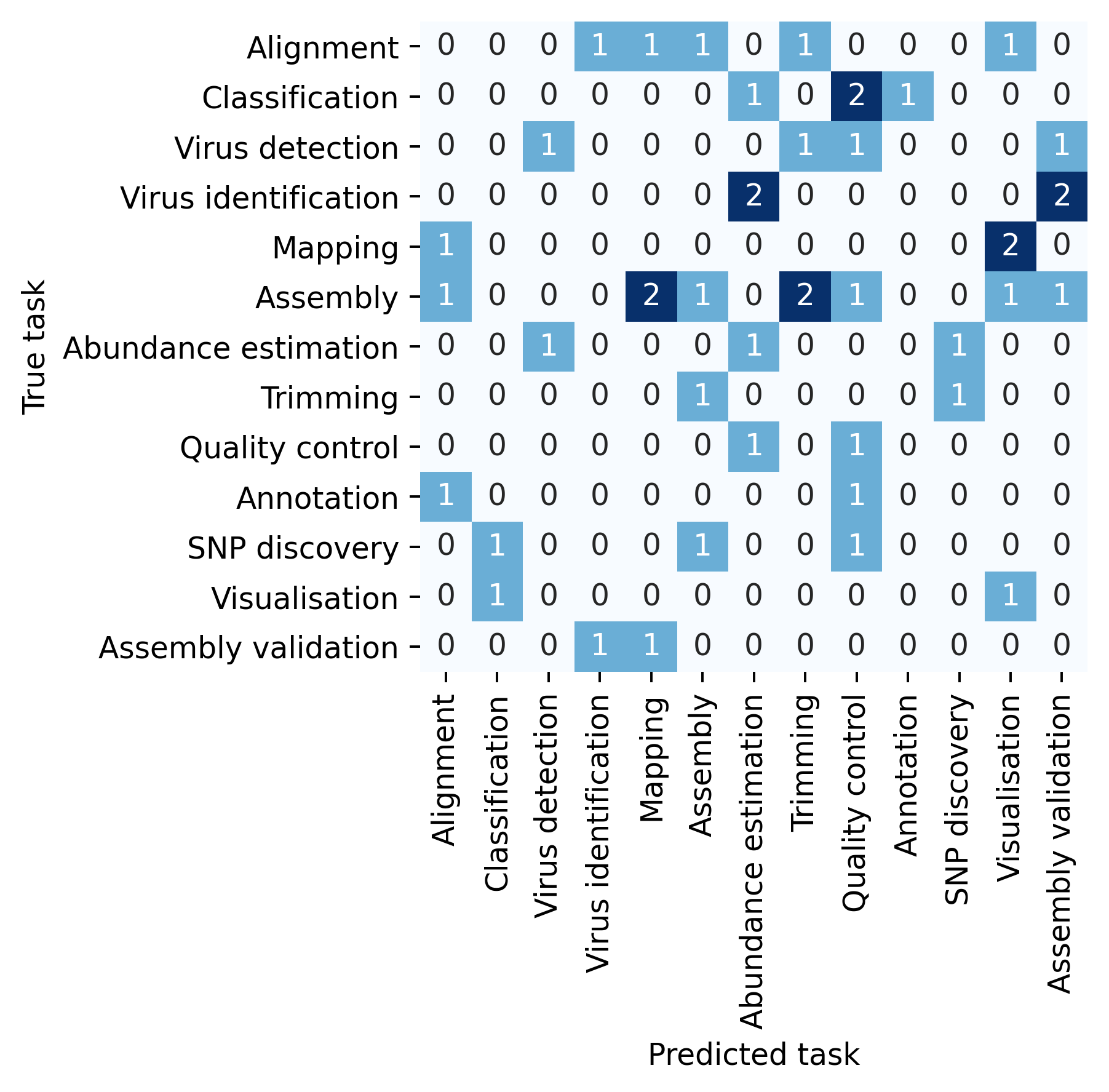}
        \caption{Logistic Regression on the ELECTRA implementation of the NLU library - test set confusion matrix - ``methods only'' dataset.}\label{sfig:LR+ELECTRA-NLU_methonly}
    \end{minipage}
\end{figure}

\clearpage

\subsection{Misclassifications done by the best model}
\subsubsection*{Description}
In Supplementary Table~\ref{stab:IndependentTestSetClassifications}, we report the eight misclassified tools by the best model (Logistic Regression on ``abstracts only'' BioBERT-hf Embeddings) from a test set of 45 examples.

\subsubsection*{Tables}

\begin{table}[!htb]
\caption{Logistic Regression on ``abstracts only'' BioBERT-hf Embeddings. Independent test set instance classifications. Misclassified examples are highlighted.}\label{stab:IndependentTestSetClassifications}
\scriptsize
\begin{tabular}{lllll}\toprule
\textbf{Tool} &\textbf{True class (TC)} &\textbf{TC probability - rank} &\textbf{Predicted class (PC)} &\textbf{PC probability } \\\midrule

GASiC &AbundanceEstimation &0.75 - 1 &AbundanceEstimation &0.75 \\
\cellcolor[HTML]{A8A8A8}dnAQET &\cellcolor[HTML]{A8A8A8}AssemblyEvaluation &\cellcolor[HTML]{A8A8A8}0.31 - 2 &\cellcolor[HTML]{A8A8A8}QualityControl &\cellcolor[HTML]{A8A8A8}0.32 \\
cutPrimers &Trimming &0.71 - 1 &Trimming &0.71 \\
GAAS &AbundanceEstimation &0.49 - 1 &AbundanceEstimation &0.49 \\
Taxonomer &Classification &0.27 - 1 &Classification &0.27 \\
\cellcolor[HTML]{A8A8A8}Kart &\cellcolor[HTML]{A8A8A8}Alignment &\cellcolor[HTML]{A8A8A8}0.21 - 3 &\cellcolor[HTML]{A8A8A8}Trimming &\cellcolor[HTML]{A8A8A8}0.3 \\
SKESA &Assembly &0.78 - 1 &Assembly &0.78 \\
Vipie &VirusIdentification &0.32 - 1 &VirusIdentification &0.32 \\
Bambino &Visualisation &0.29 - 1 &Visualisation &0.29 \\
EasyQC &QualityControl &0.59 - 1 &QualityControl &0.59 \\
VirSorter &VirusIdentification &0.25 - 1 &VirusIdentification &0.25 \\
Sailfish &AbundanceEstimation &0.63 - 1 &AbundanceEstimation &0.63 \\
AlienTrimmer &Trimming &0.41 - 1 &Trimming &0.41 \\
Bowtie &Alignment &0.62 - 1 &Alignment &0.62 \\
ATLAS-SNP2 &SNPDiscovery &0.29 - 1 &SNPDiscovery &0.29 \\
Minimus &Assembly &0.66 - 1 &Assembly &0.66 \\
Centrifuge &Classification &0.43 - 1 &Classification &0.43 \\
Kraken &Classification &0.43 - 1 &Classification &0.43 \\
ALLPATHS &Assembly &0.81 - 1 &Assembly &0.81 \\
\cellcolor[HTML]{A8A8A8}Savant &\cellcolor[HTML]{A8A8A8}Annotation &\cellcolor[HTML]{A8A8A8}0.16 - 2 &\cellcolor[HTML]{A8A8A8}Visualisation &\cellcolor[HTML]{A8A8A8}0.42 \\
cuBlASTp &Alignment &0.55 - 1 &Alignment &0.55 \\
PhageFinder &VirusDetection &0.37 - 1 &VirusDetection &0.37 \\
SOAP2 &Mapping &0.33 - 1 &Mapping &0.33 \\
Rnnotator &Assembly &0.20 - 1 &Assembly &0.2 \\
\cellcolor[HTML]{A8A8A8}consed &\cellcolor[HTML]{A8A8A8}Visualisation &\cellcolor[HTML]{A8A8A8}0.17 - 2 &\cellcolor[HTML]{A8A8A8}QualityControl &\cellcolor[HTML]{A8A8A8}0.23 \\
VirFind &VirusDetection &0.51 - 1 &VirusDetection &0.51 \\
MePIC &VirusIdentification &0.33 - 1 &VirusIdentification &0.33 \\
Minimap2 &Alignment &0.38 - 1 &Alignment &0.38 \\
SPADES &Assembly &0.57 - 1 &Assembly &0.57 \\
\cellcolor[HTML]{A8A8A8}metaSPAdes &\cellcolor[HTML]{A8A8A8}Assembly &\cellcolor[HTML]{A8A8A8}0.30 - 2 &\cellcolor[HTML]{A8A8A8}Classification &\cellcolor[HTML]{A8A8A8}0.32 \\
VERSE &VirusDetection &0.61 - 1 &VirusDetection &0.61 \\
SSPACE &Assembly &0.66 - 1 &Assembly &0.66 \\
\cellcolor[HTML]{A8A8A8}PyroBayes &\cellcolor[HTML]{A8A8A8}SNPDiscovery &\cellcolor[HTML]{A8A8A8}0.18 - 3 &\cellcolor[HTML]{A8A8A8}Mapping &\cellcolor[HTML]{A8A8A8}0.36 \\
VCAKE &Assembly &0.37 - 1 &Assembly &0.37 \\
\cellcolor[HTML]{A8A8A8}TAGdb &\cellcolor[HTML]{A8A8A8}Mapping &\cellcolor[HTML]{A8A8A8}0.05 - 7 &\cellcolor[HTML]{A8A8A8}Assembly &\cellcolor[HTML]{A8A8A8}0.37 \\
VirusDetect &VirusIdentification &0.40 - 1 &VirusIdentification &0.4 \\
Baa. pl &AssemblyEvaluation &0.19 - 1 &AssemblyEvaluation &0.19 \\
H-BLAST &Alignment &0.70 - 1 &Alignment &0.7 \\
ViraMiner &VirusIdentification &0.32 - 1 &VirusIdentification &0.32 \\
MetaShot &Classification &0.42 - 1 &Classification &0.42 \\
SSPACE-LongRead &Assembly &0.60 - 1 &Assembly &0.6 \\
CSN &Annotation &0.23 - 1 &Annotation &0.23 \\
FQC &QualityControl &0.68 - 1 &QualityControl &0.68 \\
PhiSpy &VirusDetection &0.57 - 1 &VirusDetection &0.57 \\
\cellcolor[HTML]{A8A8A8}DeepVariant &\cellcolor[HTML]{A8A8A8}SNPDiscovery &\cellcolor[HTML]{A8A8A8}0.04 - 7 &\cellcolor[HTML]{A8A8A8}Alignment &\cellcolor[HTML]{A8A8A8}0.35 \\

\bottomrule
\end{tabular}
\end{table}
\vfill

\subsubsection*{Supplementary Figures}

\begin{figure}[H]
    \centering
    \begin{minipage}{1\textwidth}
        \centering
        \fbox{\includegraphics[scale=0.34]{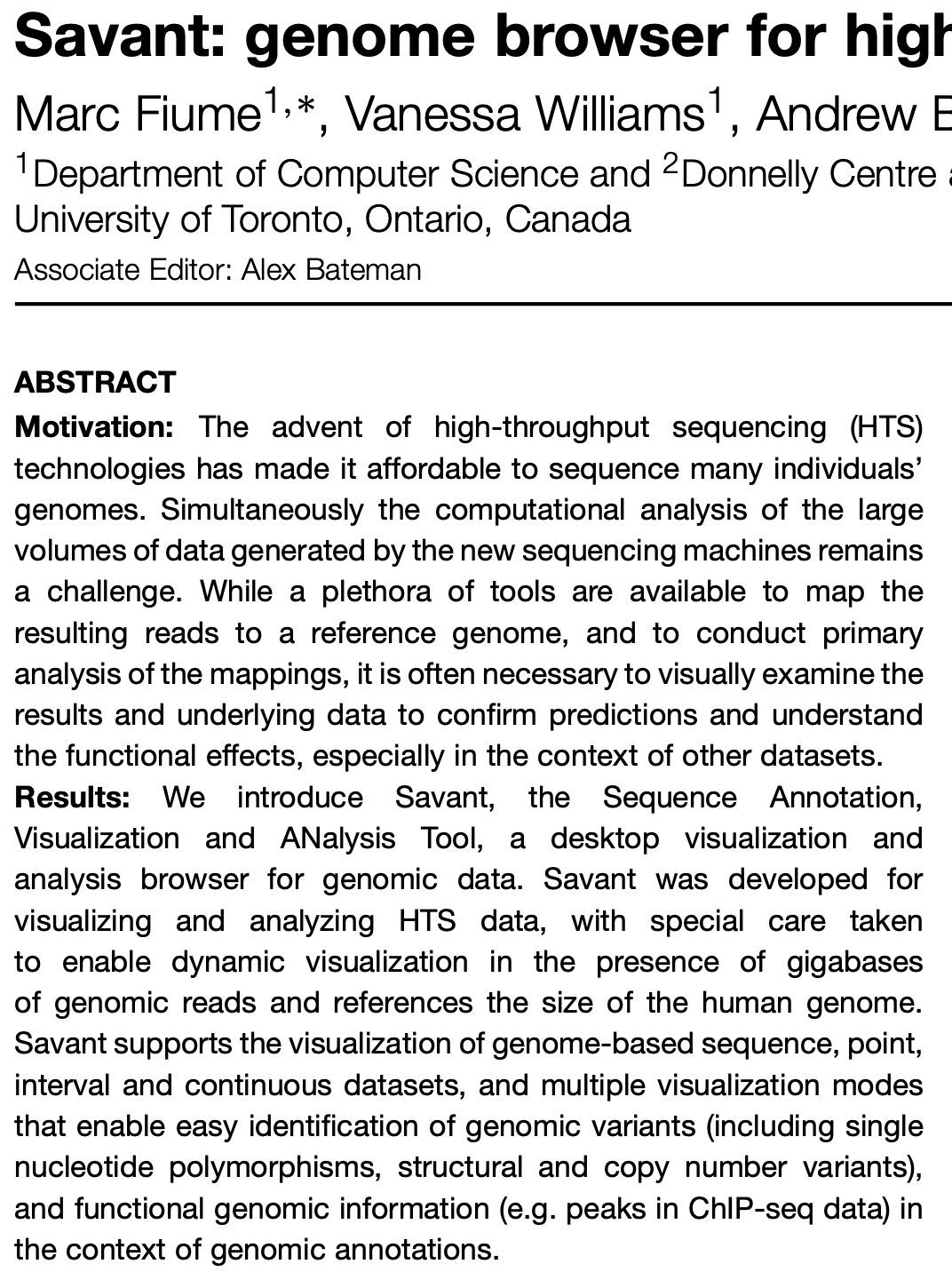}}
        \caption{Savant abstract section~\cite{Fiume2010}.}\label{sfig:Savant}
    \end{minipage}\hfill
    \bigskip{}
    \begin{minipage}{1\textwidth}
        \centering
        \fbox{\includegraphics[scale=0.34]{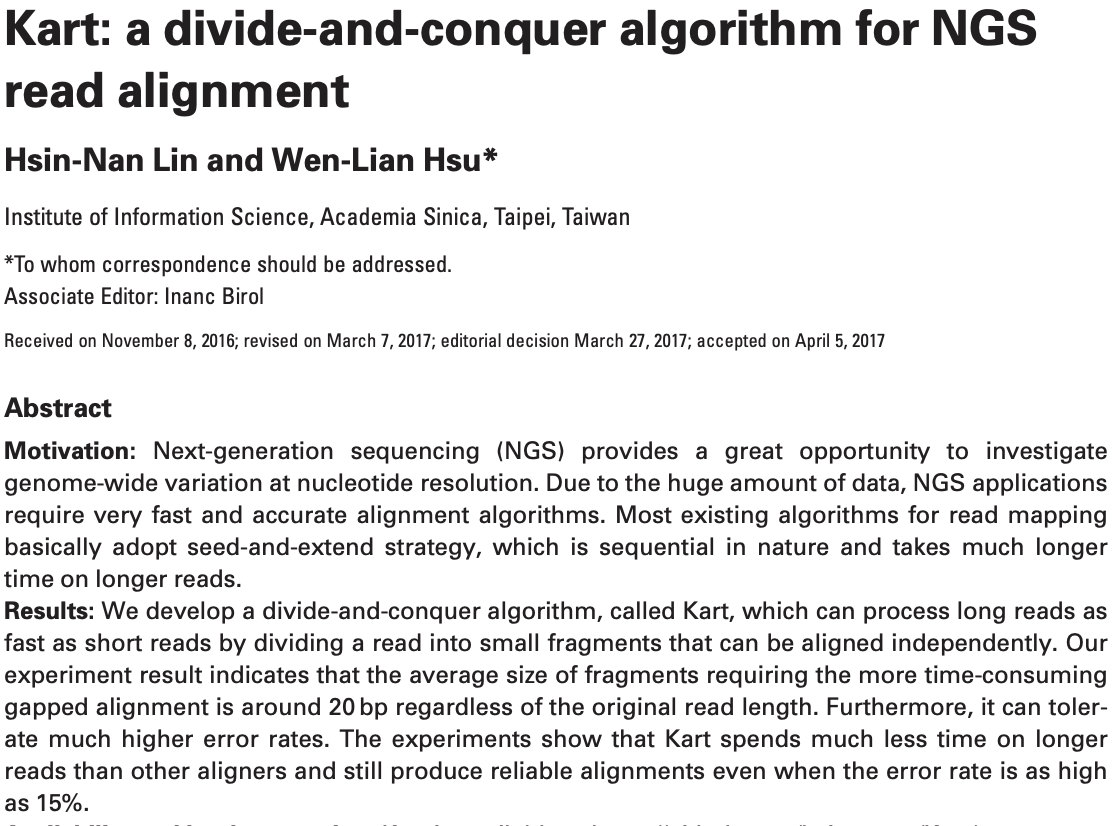}}
        \caption{Kart abstract section~\cite{Lin2017}.}\label{sfig:Kart}
    \end{minipage}\hfill
    \bigskip{}
    \begin{minipage}{1\textwidth}
        \centering
        \fbox{\includegraphics[scale=0.34]{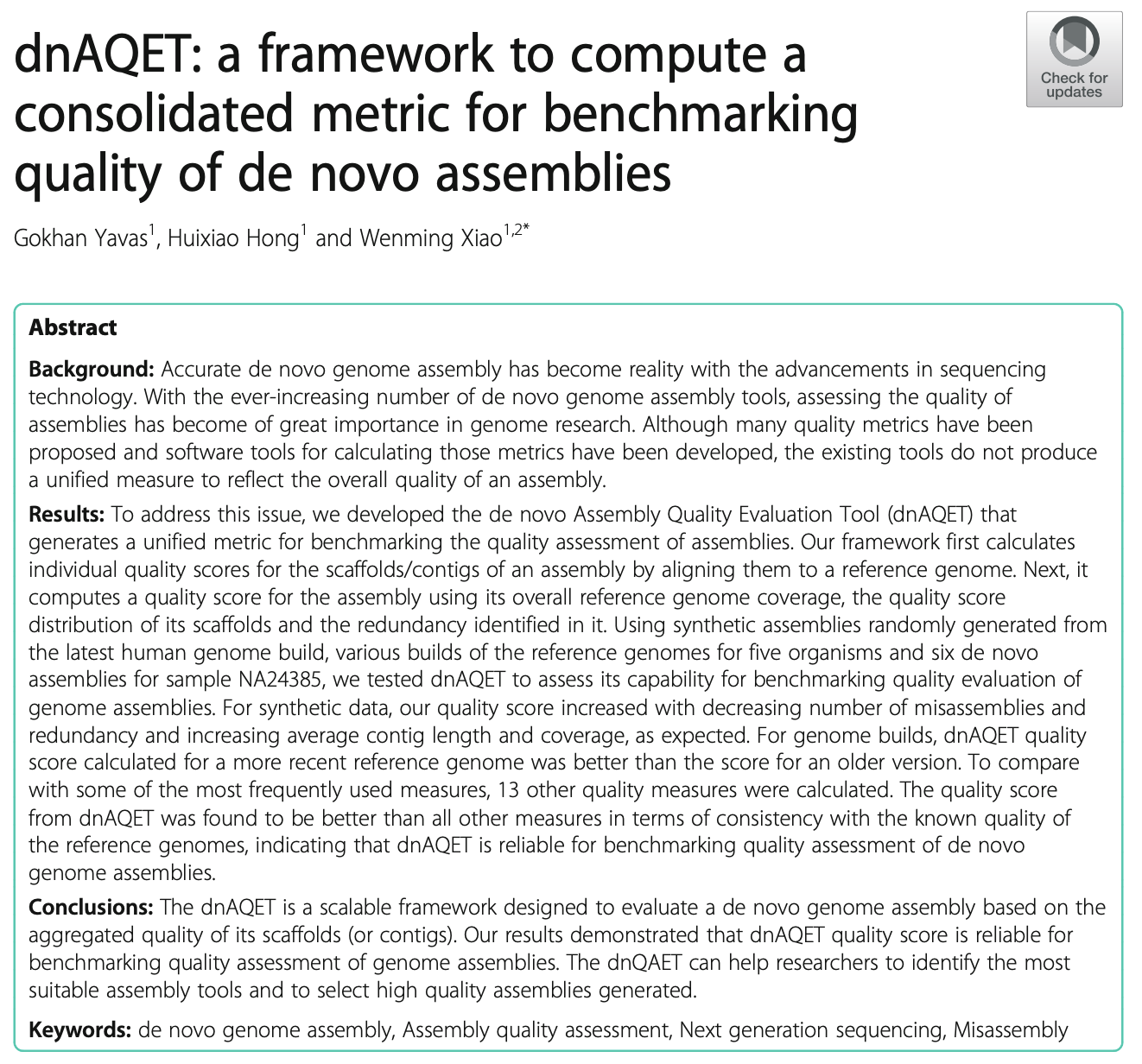}}
        \caption{dnAQET abstract section~\cite{Yavas2019}.}\label{sfig:dnAQET}
    \end{minipage}\hfill
\end{figure}
 
\end{document}